\documentclass[usenatbib]{mn2e}
\usepackage{times}
\usepackage{psfig}
\usepackage{amssymb}
\usepackage{amsbsy}
\usepackage{graphics}
\usepackage{cancel}
\input {psfig.sty}

\def \Mpc {~h^{-1}~{\rm Mpc} }

\def \Omlam {\Omega_{\Lambda}}
\def \Omm {\Omega_{\rm m}}
\def \ho {H_0}

\def \kms {{\rm ~km~s}^{-1}}
\def \kmsmpc {{\rm ~km~s}^{-1}~{\rm Mpc}^{-1}}
\def \hmpc{\;h^{-1}{\rm Mpc}}

\def \bj {b_{\rm J}}

\def \xis{\xi(s)}
\def \xisp{\xi(\sigma, \pi)}

\def \xir{\xi(r)}
\def \max {_{\rm max}}
\def \gsim { \lower .75ex \hbox{$\sim$} \llap{\raise .27ex \hbox{$>$}} }
\def \lsim { \lower .75ex \hbox{$\sim$} \llap{\raise .27ex \hbox{$<$}} }
\def \deg {^{\circ}}

\def \wrms {\langle w_{\rm z}^2\rangle^{1/2}}

\def \wp {w_{p}(\sigma)}

\title[The 2SLAQ LRG Two-point correlation function]
      {The 2dF-SDSS LRG and QSO Survey: The LRG 2-Point Correlation Function 
and Redshift-Space Distortions}
\author[N.P. Ross et al.]
       {Nicholas P. Ross$^{1}$\thanks{email: Nicholas.Ross@durham.ac.uk}, 
J. da \^{A}ngela$^{1}$, T. Shanks$^{1}$, David A. Wake$^{1}$, 
Russell D. Cannon$^{2}$, 
\newauthor A.C. Edge$^{1,3}$, R.C. Nichol$^{4}$, P.~J. Outram$^{1}$, 
Matthew Colless$^{2}$, Warrick J. Couch$^{5}$, 
\newauthor Scott M. Croom$^{2}$, Roberto De Propris,$^{6}$, Michael J. Drinkwater$^7$, Daniel J. Eisenstein$^8$, 
\newauthor Jon Loveday$^9$, Kevin A. Pimbblet$^7$, Isaac G. Roseboom$^7$, Donald P. Schneider$^{10}$, 
\newauthor Robert G. Sharp$^{3}$,   
P.~M. Weilbacher$^{11}$ \\
$^1$Physics Department, Durham University, South Road, Durham, DH1 3LE, UK.\\
$^2$Anglo-Australian Observatory, PO Box 296, Epping, NSW 1710, Australia.\\
$^3$Institute for Computational Cosmology, Durham University, South Road, 
Durham, DH1 3LE\\
$^4$Institute of Cosmology and Gravitation (ICG), University of Portsmouth,
Mercantile House, Hampshire Terrace, Portsmouth, PO1 2EG, UK.\\
$^5$Centre for Astrophysics \& Supercomputing, Swinburne University,
Hawthorn VIC 3122, Australia.\\
$^6$Cerro Tololo Inter-American Observatory, Casilla 603, La Serena, Chile\\
$^7$Department of Physics, University of Queensland, Brisbane, QLD 4072, 
Australia\\ 
$^8$Steward Observatory, University of Arizona, 933 North Cherry Avenue, 
Tucson, AZ 85721, USA\\
$^9$Astronomy Centre, University of Sussex, Falmer, Brighton BN1 9QJ\\
$^{10}$Department of Astronomy and Astrophysics, The Pennsylvania State 
University, 525 Davey Laboratory, University Park, PA 16802, USA\\
$^{11}$Astrophysikalisches Institut Potsdam, An der Sternwarte 16. 14482 
Potsdam, Germany.
}

\begin{document}

\maketitle

\begin{abstract}
We present a clustering analysis of Luminous Red Galaxies (LRGs) using
nearly \hbox{9 000} objects from the final, three year catalogue of the
2dF-SDSS LRG And QSO (2SLAQ) Survey. We measure the redshift-space
two-point correlation function, $\xi(s)$ and find that, at the mean LRG
redshift of $\bar{z} =0.55$, $\xi(s)$ shows the characteristic downturn
at  small scales ($\lsim 1\hmpc$) expected from line-of-sight velocity
dispersion.  We fit a double power-law to $\xi(s)$ and measure an
amplitude and slope of $s_0=17.3^{+2.5}_{-2.0} \hmpc$,
$\gamma=1.03\pm0.07$ at small scales  ($s < 4.5 \hmpc$) and
$s_0=9.40\pm0.19 \hmpc$, $\gamma=2.02\pm0.07$ at large scales ($s > 4.5
\hmpc$).  In the semi-projected correlation function, $w_p(\sigma)$, we
find a simple power law with $\gamma=1.83\pm0.05$ and $r_0=7.30\pm0.34
\hmpc$ fits the data in the range $0.4<\sigma<50 \hmpc$,  although
there is evidence of a steeper power-law at smaller scales.  A single
power-law also fits the deprojected correlation function $\xi(r)$, with
a correlation length of $r_{0}=7.45\pm0.35 \hmpc$ and a power-law slope
of $\gamma = 1.72\pm0.06$ in the $0.4< r <50 \hmpc$ range.  But it is
in the LRG angular correlation function that the strongest evidence for
non-power-law  features is found where a slope of
$\gamma=-2.17\pm{0.07}$  is seen at $1<r<10 \hmpc$ with a flatter
$\gamma=-1.67\pm{0.07}$ slope  apparent at $r \, \lsim \, 1 \hmpc$
scales.

We use the simple power-law fit to the galaxy $\xi(r)$, under the
assumption of linear bias, to model the redshift space distortions in
the 2-D redshift-space correlation function, $\xi(\sigma,\pi)$. We fit
for the LRG velocity dispersion, $w_z$, the density parameter, $\Omm$
and $\beta(z)$, where $\beta(z)=\Omm^{0.6}/b$ and $b$ is the linear
bias parameter. We find values of $w_z=330$kms$^{-1}$, $\Omega_{\rm m}
= 0.10^{+0.35}_{-0.10}$ and $\beta = 0.40\pm0.05$. The low values for
$w_z$ and $\beta$ reflect the high bias of the LRG sample. These high
redshift results, which incorporate the Alcock-Paczynski effect and the
effects of dynamical infall, start to break the degeneracy between
$\Omm$ and $\beta$ found in low-redshift galaxy surveys such as 2dFGRS.
This degeneracy is further broken by introducing an additional external
constraint, which is the value $\beta(z=0.1)=0.45$ from 2dFGRS, and
then considering the evolution of clustering from $z \sim 0$ to $z_{\rm
LRG} \sim 0.55$. With these combined methods we find $\Omega_{\rm
m}(z=0) = 0.30\pm0.15$ and $\beta(z=0.55)= 0.45\pm0.05$. Assuming these
values, we find a value for $b(z=0.55)=1.66\pm0.35$. We show that this
is consistent with a simple ``high-peaks'' bias prescription which
assumes that LRGs have a  constant co-moving density and their
clustering evolves purely under gravity.
\end{abstract}

\begin{keywords}
galaxies: clustering -- luminous red galaxies: general -- cosmology: 
observations -- large-scale structure of Universe.
\end{keywords}

%%%%%%%%%%%%%%%%%%%%%%%%%%%%%%%%%%%%%%%%%%%%%%%%%%%%%%%%%%%%%%%%%%%%
%SECTION 1  SECTION 1  SECTION 1  SECTION 1  SECTION 1  SECTION 1  %
%SECTION 1  SECTION 1  SECTION 1  SECTION 1  SECTION 1  SECTION 1  %
%SECTION 1  SECTION 1  SECTION 1  SECTION 1  SECTION 1  SECTION 1  %
%%%%%%%%%%%%%%%%%%%%%%%%%%%%%%%%%%%%%%%%%%%%%%%%%%%%%%%%%%%%%%%%%%%%
\section{Introduction}

Recent measurements of the galaxy correlation function, $\xi$, have
produced a series of impressive results. Whether it be the detection of
baryonic acoustic oscillations \citep{Eisenstein05}, clustering
properties of different spectral types of galaxy \citep{Madgwick03}, or
the evolution of AGN black hole mass \citep{Croom05}, the two-point
correlation function continues to be a key statistic when studying
galaxy clustering and evolution. There have also been a series of recent
studies \citep[e.g.,][]{Zehavi05a, LeFevre05, Coil04, Phleps06}
investigating the clustering properties and evolution
with redshift of galaxies from 0.3 $< z <$ 1.5. Amongst these,
\citet{Zehavi05a} use the Sloan Digital Sky Survey \citep[SDSS;][]{York00} 
to examine the clustering properties of Luminous Red Galaxies (LRGs) 
at a redshift of z$\simeq$0.35. 
They find that correlation length depends on LRG
luminosity and that there is a deviation from a power-law in the
real-space correlation function, with a dip at $\sim$ 2 Mpc scales as
well as an upturn on smaller scales.

Although the form of  the 2-point correlation function is in itself a
worthwhile cosmological datum, more information can be gained by
studying the dynamical distortions at both small and large
scales in the clustering pattern \citep{Kaiser87}. 
Measured galaxy redshifts consist of a component from the 
Hubble expansion plus the motion induced by the galaxy's local potential. 
This leads to one type of distortion in {\it redshift}-space 
from the {\it real}-space clustering pattern. 
There are two basic forms of dynamical distortion (a) small scale 
virialised velocities causing elongations in redshift direction - 
`Fingers of God', but at larger scales there will also be flattening 
of the clustering in the redshift direction due to dynamical infall. 
Another type of geometric distortion can be introduced if we assume 
the wrong cosmology  to convert redshifts to comoving distances \citep{AP79}. 
Under the assumption that galaxy clustering is isotropic in real-space, 
a test can be performed in redshift-space by determining which 
cosmological parameters return an isotropic clustering pattern.

In the linear regime, dynamical effects are broadly determined by the parameter
$\beta$, where $\beta=\Omega_{\rm m}^{0.6} / b$, $\Omm$ is the matter density 
parameter and $b$ is the linear bias 
parameter. If we assume, as is common, a zero spatial curvature
model, then the main parameter determining geometric distortion is
$\Omm$. We can  therefore use these redshift-space distortions to
our advantage and derive from them estimates of $\Omega_{\rm m}$ and $\beta$, 
\citep[e.g.,][]{Kaiser87, Loveday96, Matsubara96, Matsubara01, Ballinger96,
Peacock01, Hoyle02, daAngela05}. Unfortunately,
there is often a degeneracy between these parameters, but this can be
broken by the inclusion of other information. This additional information is
introduced via constraints obtained from linear evolution
theory of cosmological density perturbations \citep[and references therein]
{daAngela05}.

In this work, we extend the redshift coverage of the SDSS LRG survey by
using the data from the recently completed 2dF-SDSS LRG And QSO (2SLAQ) Survey
(Cannon et al. (2006); Croom et al. (2007), in prep.).
Luminous Red Galaxies are ideal candidates for galaxy redshift 
surveys since they are intrinsically bright and so can be seen to 
cosmological distances. Selection criteria are used which gave a 
relatively clean and complete selection of LRGs and since 
they are the most massive galaxies, they are believed to reside 
in over-dense peaks of the underlying matter distribution and are thus 
excellent tracers of large scale-structure. 

Observations of the 2SLAQ Survey are now complete, with a number of new
results being reported (e.g. \citet{Wake06}, \citet{Roseboom06},
Sadler et al. (2007), in prep.). 
In this paper we shall concentrate on the clustering
of the 2SLAQ LRG sample, extending the work of the SDSS LRG Survey
\citep{Eisenstein01, Zehavi05a} to higher redshift. We
calculate the 2-point galaxy correlation function in both redshift-space
and real-space for LRGs over the redshift range $0.4 < z < 0.8$. 
Then using information gained from geometric distortions in the
redshift-space clustering pattern, values of the cosmological parameters
$\Omega_{\rm m}$ and $\beta$ can be found \citep[e.g.][]{AP79, Ballinger96, 
Hoyle02, daAngela05}.

In Section 2 we therefore introduce the 2SLAQ sample and the techniques
used in our analysis. In Section 3 the 2SLAQ LRG correlation function
measurements are presented and comparisons to other surveys are made. In
Section 4 we model the redshift-space distortions and compare these
models to our data, finding values of $\Omega_{\rm m}$ and $\beta$. Our
conclusions are presented in Section 5.

%%%%%%%%%%%%%%%%%%%%%%%%%%%%%%%%%%%%%%%%%%%%%%%%%%%%%%%%%%%%%%%%%%%%
%SECTION 2  SECTION 2  SECTION 2  SECTION 2  SECTION 2  SECTION 2  %
%SECTION 2  SECTION 2  SECTION 2  SECTION 2  SECTION 2  SECTION 2  %
%SECTION 2  SECTION 2  SECTION 2  SECTION 2  SECTION 2  SECTION 2  %
%%%%%%%%%%%%%%%%%%%%%%%%%%%%%%%%%%%%%%%%%%%%%%%%%%%%%%%%%%%%%%%%%%%%
\section{Data and Techniques}\label{section_data} 
\begin{figure*}
\centering
\centerline{\psfig{file=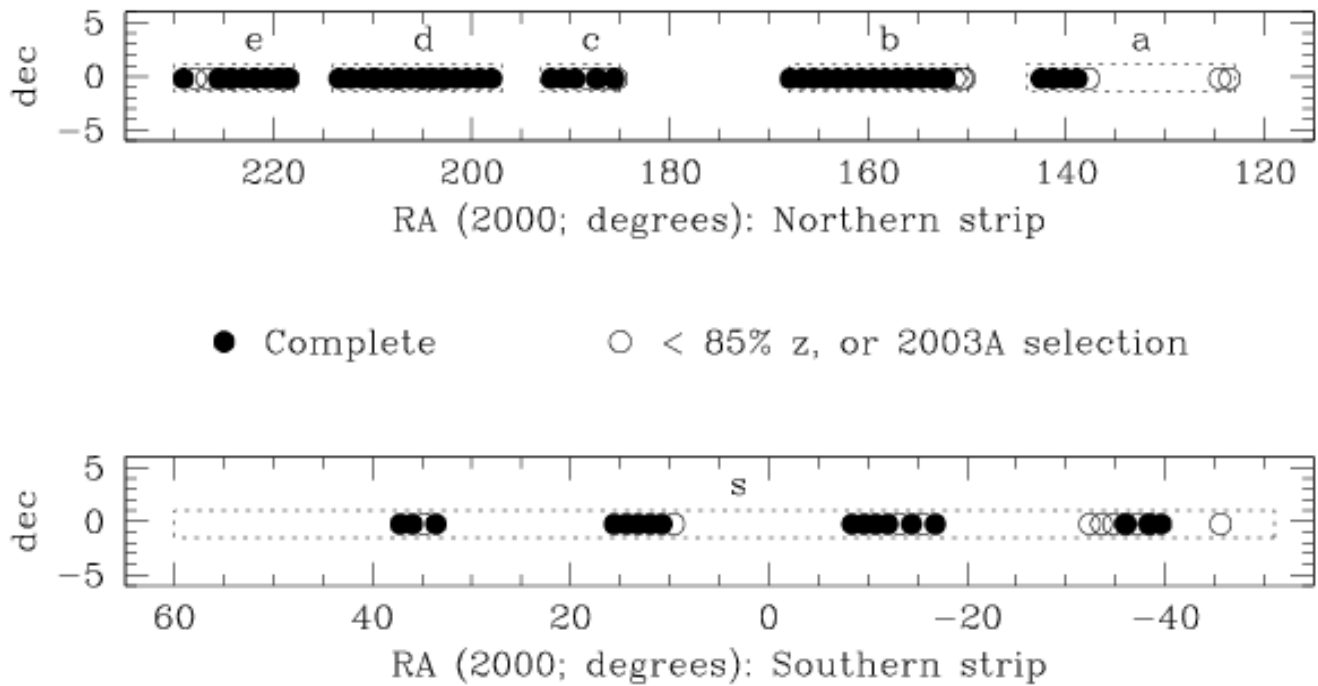,width=18.0cm}}
\caption{The location of the 2SLAQ Input Catalogue (dotted rectangles) and 
observered fields (circles). Solid circles indicate fully observed fields with
high, $\geq 85\%$ overall completeness, 
while hollow circles have less than 85\% overall completeness 
or fields with non-standard selection criteria.} 
\label{fig:stripes}
\end{figure*}

\subsection{The 2dF-SDSS LRG And QSO Survey} \label{sec:2SLAQ_details}

A full description of the 2SLAQ Survey can be found in
\citet{Cannon06}.  At its heart, the 2SLAQ Survey relies on the SDSS
photometric survey to supply LRG targets for spectroscopic follow-up
using the  2 Degree Field (2dF) instrument on the Anglo-Australian
Telescope (AAT).

The selection of distant ($z >$ 0.4) LRGs is done on the basis of SDSS
$gri$  photometric data, using the ($g-r$) versus ($r-i$) colours and
the SDSS ``de Vaucouleurs'' $i$-band magnitude.  The criteria are
similar to those used for the faint ``Cut II" sample in the SDSS LRGs
\citep{Eisenstein01}  and are described in detail by
\citet{Cannon06}. (See \citet{Fukugita96} for a description of the SDSS
filters.)

The survey covers two narrow stripes along the celestial equator
($\vert$$\delta$$\vert$ $<$ 1.5$\deg$).  The Northern Stripe runs from
$8\fh4$ to $15\fh3$ in Right Ascension and  is broken into 5
sub-stripes to utilise the best photometric data.  The Southern Stripe
runs from $20\fh6$ to $4\fh0$. Figure \ref{fig:stripes} shows the
layout of the target stripes and the 2dF fields observed.  The total
area of the survey, including the overlap regions,  was approximately
180 degrees$^2$.  Again, complete details of the Survey fields are
given by \citet{Cannon06}.

It is important to be aware of the tiling strategy  of the 2SLAQ survey
when estimating the clustering of the LRGs.  A simpler tiling scheme
was used for 2SLAQ than for the preceding  2dFGRS/2QZ survey. For
instance, for 2SLAQ, the 2dF tiles were offset by  1.2 deg in the RA
direction as opposed to a variable spacing strategy employed by the
2dFGRS and 2QZ.  Again, contrary to the 2dFGRS/2QZ, the galaxies in
2SLAQ were given higher  fibre assignment priority, with the LRGs
always having priority over the QSOs.  This makes sure the LRG
selection was not biased by the QSOs.  The details of the survey mask
and selection function will be described  in detail in Section
\ref{sec:themask}.

The total 2SLAQ LRG dataset consists of a total of \hbox{18 487}
spectra for \hbox{14 978} discrete objects;  \hbox{13 784} of these
(92\%) have reliable, ``Qop'' $ \geq 3$ redshifts.\footnote{``Qop''
represents a  redshift quality flag assigned by visual inspection of
the  galaxy spectrum and the redshift cross-correlation function. A
value of 3 or greater represents a 95-99\% confidence that the redshift
obtained from the spectrum is valid.}  From these ``Qop''$ \geq 3$
objects, 663 are identified as being stars, leaving a total of \hbox{13
121} galaxies.

We cut this sample down further by using only those confirmed LRGs
which were part of the top priority ``Sample 8'' selection as described
fully in \citet{Cannon06}. These galaxies comprise the most rigorously
defined 2SLAQ LRG sample where completeness is highest due to their top
priority for spectroscopic observation. The exact Sample 8 selection
lines in the $gri$ plane are shown in Fig. 1 of Cannon et al (2006).
The magnitude limits is $i_{deV}<$ 19.8 (de-reddened).  However, the
sample we use does include observations taken in the 2003A semester,
where a brighter $i_{deV}<$ 19.5 magnitude limit was used, as long as
the observed LRG would have made the ``Sample 8'' selection. We do not
include observations taken from fields a01, a02 and s01 (see
\citet{Cannon06}) as they have low completeness and should not be used
in statistical analyses.  Once the final selection criteria had been
decided, there were \hbox{25 795} ``Sample 8'' LRG targets at a sky
density of about 70 per square degree. Approximately 40\% (\hbox{10
072}) of these objects  were observed, with \hbox{9 307} obtaining
``Qop''$ \geq 3$.  After imposing the cuts above, this leaves a total
of  \hbox{8 656} LRGs, \hbox{5 995} in the Northern Galactic Stripe and
\hbox{2 661} in the Southern Galactic Stripe. For all further analysis,
this is the sample utilised which we call the ``Gold Sample'' and has a
$\bar{z}_{Gold}$ = 0.55.

\begin{table}
\begin{center}
\setlength{\tabcolsep}{4pt}
\begin{tabular}{lrrr}
\hline
\hline
Sample Description  & Number in sample & North & South   \\
\hline
Unique Objects               & 14 978 & 10 369 & 4 609   \\
``Qop'' $\geq$ 3             & 13 784 &  9 726 & 4 058   \\
M Stars                      &    663 &        &         \\
LRGs                         & 13 121 &  9 280 & 3 841   \\ 
LRG Sample 8                 &  8 756 &  6 076 & 2 680   \\
\,\,\,  excl. a01, a02, s01  & {\bf 8 656} & {\bf 5 995} & {\bf 2 661}  \\
\hline
\hline
\label{tab:The_LRG_numbers}
\end{tabular}
\caption{The 2SLAQ LRG Survey; Numbers of different Samples. 
Over \hbox{18 000} spectra were obtained, resulting in 
\hbox{13 121} spectroscopically confirmed
Luminous Red Galaxies. We use the LRGs with the ``Sample 8''
Input Priority settings for our analysis but do not include the data 
taken in the a01, a02 and s01 fields which have low redshift completeness 
and should be excluded from statistical analysis \citep{Cannon06}. 
Thus we are left with \hbox{8 656} in our ``Gold Sample''. }
\end{center}
\end{table}

\subsection{The Two-Point Correlation Function} \label{sec:errors}
Here we give a brief description of the 2-point correlation function (2PCF); 
for a more formal treatment the reader is referred to \citet{Peebles80} 
which presents the basis for the rest of the section. 
To denote the {\it redshift}-space 
(or $z$-space) correlation function, we will use the notation $\xi(s)$ 
and to denote the {\it real}-space correlation function, $\xi(r)$ will be used,
where $s$ is the redshift-space separation of two galaxies and 
$r$ is the real-space separation. 

The 2-point correlation function, $\xi(x)$, 
is defined by the joint probability that 
two galaxies are found in the two volume elements $dV_{1}$ and 
$dV_{2}$ placed at separation $x$,
\begin{equation}
	\label{equ:19point2}
	dP_{12} = n^{2} [1 + \xi(x)] \, dV_{1} dV_{2}.
\end{equation}
To calculate $\xi(x)$, $N$ points are given inside a window $W$ 
of observation, which is a three-dimensional body of volume $V(W)$. 
An estimation of $\xi(x)$ is based on an average of the 
counts of neighbours of galaxies at a given scale, or more precisely, 
within a narrow interval of scales. 
An extensively used estimator is that of \citet{Davis83} 
and is usually called the standard estimator,
\begin{equation}
  \xi_{Std}(s) = \left( \frac{N_{rd}}{N}  \frac{DD(s)}{DR(s)} \right) - 1
\end{equation}
where $DD(s)$ is the number of pairs in a given catalogue 
(within the window $W$) and $DR(s)$ is the number of pairs 
between the data and the random sample with separation in the same interval.
$N_{rd}$ is the total number of random points and $N$ is the total 
number of data points.
A value of $\xi$ = 1 implies there are twice as many pairs of galaxies 
than expected for a random distribution and the scale at which this is 
the case is called the correlation length.

\subsection{Constructing a random catalogue and survey 
Completeness}\label{sec:themask}
The two point correlation function, $\xi$, is measured  by comparing
the actual galaxy distribution to a catalogue of randomly  distributed
galaxies. Following the method of \citet{Hawkins03} and
\citet{Ratcliffe98c}, these randomly distributed galaxies are subject
to the same redshift, magnitude and mask constraints as the real data
and we modulate the surface density of points in the random catalogue
to follow the completeness variations. We now look at the various
factors  this involves.

Following \citet{Croom04}, we discuss issues regarding the 2SLAQ Survey
completeness.  As with the rest of the paper, we are only concentrating
on the properties of the luminous red galaxies.  One might think the
parallel 2SLAQ QSO survey would have a bearing on  subsequent
discussion but due to the higher priority given to the fibres  assigned
to observe the LRGs, the QSO Survey has no impact on LRG clustering
considerations, as already noted. For more description of the
clustering of  the QSOs the reader is referred to \citet{daAngela06}. 

Three main, separate types of completeness are going to be considered;
i) Coverage completeness, $f_c$, which  we define as the fraction of
the Input 2SLAQ catalogue sources that have spectroscopic
observations. Identically to \citet{Croom04}, we calculate $f_c$, as
being the ratio of observed to total sources in each of the
sectors defined by overlapping 2SLAQ fields, which are pixelized on 1
(one)  arcminute scales; ii) Spectroscopic completeness, $f_s$  which
can be said to be the fraction of observed objects which have a certain
spectroscopic quality; iii) Incompleteness due to fibre collisions
which is dealt with separately from coverage completeness

For coverage completeness and spectroscopic completeness we assume that
both are functions of angular position only, i.e. $f_c(\theta)$ and
$f_s(\theta)$ respectively. The spectroscopic (i.e. redshift)
completeness does depend  on magnitude but this is not relevant for any
of the purposes of this paper.

\subsubsection{Angular $+$ Spectroscopic Completeness and Fibre
Collisions} 
There are various technical details associated with the 2dF instrument.
Variations in target density, the small number of broken or otherwise
unuseable fibres and constraints owing to the minimum fibre placing
(see below) could introduce false signal into the
clustering pattern. For our analysis, the 2SLAQ survey consists of 80
field pointings. Many of these pointings overlap, alleviating
some of these technical issues. 

The design of the 2dF instrument means that fibres cannot 
be placed closer than approximately 30 arcsec \citep{Lewis02} so both 
members  of a close pair of galaxies cannot be targeted in a single 
fibre configuration. The simple, fixed-spacing tiling strategy of the 2SLAQ
Survey means that not all such close pairs are lost. Neighbouring tiles have
significant areas of overlap and much of the survey sky area is
targeted more than once. This allows us to target both galaxies in some
close pairs. Nevertheless, the survey misses a noticeable fraction of
close pairs. It is important to assess the impact of this omission on
the measurement of galaxy clustering and to investigate schemes that can
compensate for the loss of close pairs.

\begin{figure}
\psfig{figure=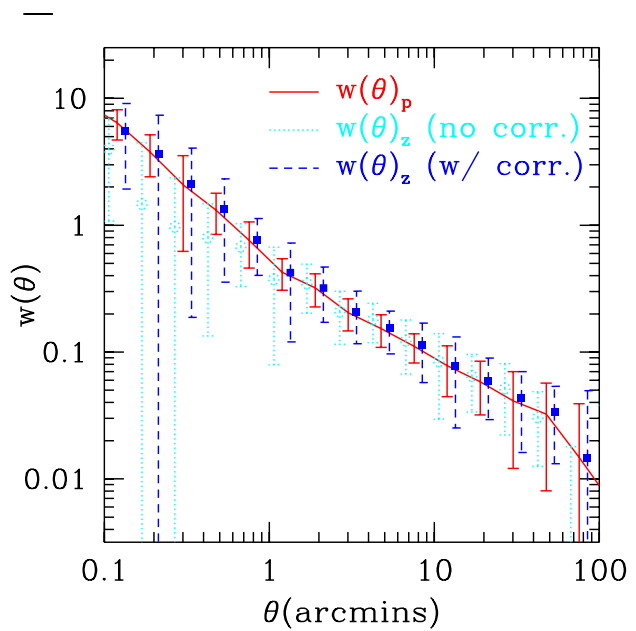,angle=0,width=8.3cm}
\centering
\caption[$w(\theta)$ for the 2SLAQ redshift catalogue] 
{The $w(\theta)$ for the 2SLAQ redshift catalogue (light blue) dotted,
open  circles compared to the parent catalogue solid (red) line.  The
errors quoted are ``Field-to-field'' errors with the sub-areas used
given by Table~\ref{tab:The_Fields_for_errors}.  The filled blue
squares, with dashed error bars,  show the $w(\theta)$ from the
redshift catalogue after the correction for fibre collisions has been
applied.   The values for the uncorrected (corrected) $w(\theta)$ from
the  redshift catalogue have been moved by $\Delta \log = -(+)0.05$  in
the abscissa for clarity.   Note also that the solid line represents 
the filled squares given in Figure~\ref{fig:w_theta_2SLAQ_labels}.}
\label{fig:w_theta}
\end{figure}

To quantify the effect of these so-called `fibre collisions' we have
followed previous 2dF studies \citep[e.g.][]{Hawkins03, Croom04}  and
calculated the angular correlation function for galaxies in the 2SLAQ
parent catalogue, $w_p(\theta)$, and for galaxies with redshifts used
in our $\xi$ analysis, $w_z(\theta)$. We used the same mask to
determine the angular selection for each sample.

As shown in Figure~\ref{fig:w_theta}, on scales $\theta \; \gsim \;
3'$,  the angular correlations of the Parent and Redshift catalogue are
very nearly consistent.  At scales $\theta \; \lsim \; 2'$, we begin to
lose close pairs.  To correct for this effect, we use a similar method
to \citet{Hawkins03} and \citet{Li06a}. The quantity $w_{\rm
cor}(\theta) =  (1 + w_{p})/(1 + w_{z})$ is used to weight our 3-D DD
pairs.  For each DD pair, the angular separation on the sky is
calculated and  the galaxy-galaxy pair is weighted by the $w_{\rm
cor}(\theta)$ ratio given by the relevant angular separation.  The
result of weighting by this factor, is shown by the filled (dark blue)
squares in Fig.~\ref{fig:w_theta}.

The last stage in determining the angular ``mask'' is to evaluate the
spectroscopic completeness of the survey, $f_s(\theta)$ which for our
purposes, we again assume depends on sky position only. This function
essentially describes the success rate in obtaining a spectrum  and
reliable redshift for a given fibred object. Here the advantage of LRGs
becomes apparent. With their well-defined early-type spectra and often
very strong Ca H+K break around 4000\AA, a high success rate was
achieved when calculating a redshift for  the 2SLAQ LRG objects. Also,
it became apparent that our 4 hour  per field exposure time was on
occasion generous and relatively high S/N spectra were recorded. The
spectroscopic completeness has been estimated at 94.5 per cent for the
primary ``Sample 8'' and the redshift completeness at 96.7 per cent,
giving an overall completeness  of 91.4 per cent  (Cannon  et al. 2006,
Section 5.5, Figure 5).

\subsubsection{Radial Selection Function and Estimates of the 
LRG $N(z)$}\label{sec:nzest}
\begin{figure}
\centering
\centerline{\psfig{file=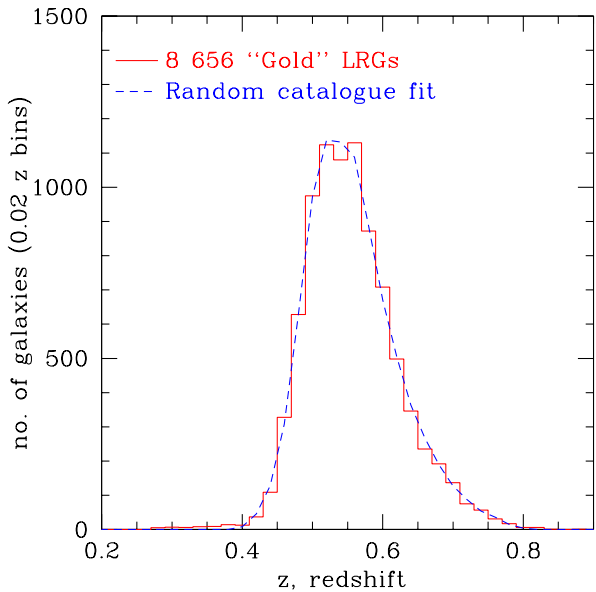,width=8cm}}
\caption{The redshift distribution for the 2SLAQ LRG ``Gold'' Sample we use.
The solid red histogram is for the ``Gold'' Sample. 
The dashed blue line is from the normalised random catalogue.}
\label{fig:nz}
\end{figure}

The observed distribution of galaxy redshifts is given in Figure \ref{fig:nz}.
Plotted are the $N(z)$ distributions, binned into redshift slices
of $\Delta z$=0.02, for the ``Gold Sample''.
Also shown is a polynomial fit (7th order) to the $N(z)$ distribution, 
which is used to generate the random distributions.  
Checking the $N(z)$ fits using higher order polynomials or convolved double 
Gaussians does not give tighter reproduction of the observed LRG redshift
distribution.

Combining the radial selection function and the completeness map, 
we generate a random catalogue of points which 
we now use to calculate the LRG correlation function.

\subsection{Calculating the 2-point Correlation Function}
As the LRG correlation function, $\xi(s)$, probes high redshifts and
large scales, the measured values are highly dependent on the assumed
cosmology.  In determining the comoving separation of pairs of
LRGs we choose to calculate $\xi(s)$ for two representative cosmological 
models.  The first uses the cosmological parameters derived 
from WMAP, 2dFGRS and other data 
\citep{Spergel03, Spergel06, Percival02, Cole05, Sanchez06} with 
$(\Omm$,$\Omlam)=$ $(0.3,0.7)$, which we will call the $\Lambda$ cosmology.  
The second model assumed is an Einstein-de Sitter 
cosmology with $(\Omm$,$\Omlam)=$ $(1.0,0.0)$
which we denote as the EdS cosmology.  We will quote distances in
terms of $\Mpc$, where $h$ is the dimensionless Hubble constant such
that $\ho=100h\kmsmpc$.

We have used the minimum variance estimator suggested by \citet{LS93}
to calculate $\xi(s)$. Using notation from 
\citet{Martinez02book}, this estimator is
\begin{eqnarray}
 \xi_\mathrm{LS}(s) &=& 1+\left( \frac{N_{rd}}{N} \right)^{2} 
                                 \frac{DD(s)}{RR(s)} - 
                 2   \left( \frac{N_{rd}}{N} \right) \frac{DR(s)}{RR(s)} \\
	     &\equiv&  \frac{\langle DD \rangle - \langle 2DR \rangle + 
                             \langle RR \rangle}
                            {\langle RR \rangle},
\label{lseq}
\end{eqnarray}
where the angle brackets denote the suitably normalised 
LRG-LRG, LRG-random and random-random pairs counted at separation $s$.
We use bin widths of $\delta\log(s / \hmpc)$ = 0.1.
The density of random points used was 20 times the density of LRGs. 
The Hamilton estimator is also utilised \citep{Hamilton93} where  
\begin{eqnarray}
  \xi_\mathrm{Ham}(s) =  \frac{DD(s) \cdot RR(s)}{DR(s)^{2}} - 1
\label{hameq}
\end{eqnarray}
and no normalisation is required. 
Since we find the differences of the Hamilton estimator 
compared to the Landy-Szalay method are negligible, 
the Landy-Szalay method is quoted in all $\xi(s)$
figures unless explicitly stated otherwise. 

Three methods are employed to estimate the likely
errors on our measurements. 
The first is a calculation of the error on $\xi(s)$ using the 
Poisson estimate of
\begin{equation}
\sigma_\mathrm{Poi}(s)=\frac{1+\xi(s)}{\sqrt{DD(s)}}.
\label{equ:xierr_Poisson}
\end{equation}

The second error estimate method is what we shall call the
{\it field-to-field} errors, calculated by 
\begin{equation}
\sigma_\mathrm{FtF}^{2}(s)=\frac{1}{N-1} \sum_{i=1}^{N}
                           \frac{DR_{i}(s)}{{DR(s)}}  [\xi_{i}(s) - \xi(s)]^{2}
\label{equ:xierr_FtF}
\end{equation}
where $N$ is the total number of subsamples i.e. ``the fields'' and
$\xi_{i}(s)$ is from one field. $\xi(s)$ is the value for $\xi$ from the 
entire sample and is not the mean of the subsamples. 
For our studies the natural unit of the ``Field-to-field'' (FtF) subsample is
given by the area geometry covered by the survey. Thus we take $N=9$, 
and split the NGP area into five regions, {\it a,b,c,d,e} and 
the SGP in to four regions, named {\it s06, s25, s48, s67}. 
Details of the FtF subsamples are given in 
Table~\ref{tab:The_Fields_for_errors}.

The third method is usually referred to as the {\it jackknife} estimate, 
and has been used in other correlation studies 
\citep[e.g.][]{Scranton02, Zehavi02, Zehavi05a}. 
Here we estimate $\sigma$ as 
\begin{equation}
\sigma_\mathrm{Jack}^{2}(s)=\sum_{i^{\prime}=1}^{N} 
                            \frac{DR_{i^{\prime}}(s)}{{DR(s)}}
                             [\xi_{i^{\prime}}(s) - \xi(s)]^{2}
\label{equ:xierr_jack}
\end{equation}
where $i^{\prime}$ is used to signify the fact that each time we  
calculate a value of $\xi(s)$, all subsamples are used bar one. 
For the jackknife errors, we divide the survey into 32 approximately 
equal sized areas, leaving out $\sim$4.5 square degrees from the entire survey
area at one time. Thus a jackknife subsample will contain $\sim$8,350 LRGs. 
We can then work out the covariance matrix in the traditional way, 
\begin{equation}
        {\rm Cov}(\xi_{i},\xi_{j}) = \frac{N-1}{N} \,
                                     \sum_{l=1}^{N} \,
                            (\xi_{i}^{l} - \bar{\xi}_{i}^{l}) \,
                            (\xi_{j}^{l} - \bar{\xi}_{j}^{l})
\end{equation}
where $\bar{\xi}$ is the mean value of $\xi$ measured from all the jackknife 
subsamples and $N=32$ in our case (c.f. \citet{Zehavi02}).
The variances are obtained from the leading diagonal elements of the 
covariance matrix, 
\begin{equation}
\sigma_{i}^{2} =  {\rm Cov}(\xi_{i},\xi_{i})
\label{eq:JK2}
\end{equation}
When examing the covariance matrix, we find the measurements to be
slightly noisy as well as an indication of anti-correlation
(contrary to theoretical expectations). However, we note that in
the other recent clustering studies, noisy covariances and
anti-correlations were also noted 
\citep[e.g.][]{Scranton02, Zehavi02, Zehavi05a}.

\begin{figure}
\centering
\centerline{\psfig{file=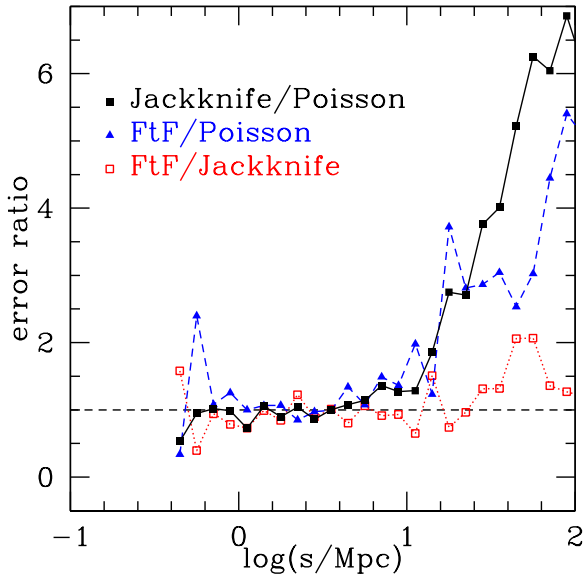,width=8cm}}
\caption[The error ratio for Poisson, Field-to-field and jackknife
errors] {The ratio of Poisson to jackknife errors (solid black
line and squares),   Poisson to `field-to-field' errors,  (dashed blue
line and triangles) and  `field-to-field' to jackknife errors (dotted
red line and open squares).  As can be seen, all error estimators are
comparable on scales $ \lesssim 10 \hmpc $, while at larger scales than
this the jackknife and `field-to-field' errors are considerably larger
than the simple Poisson estimates.  The magnitude of the
`field-to-field' and jackknife errors are very similar from the
smallest scales considered here up to  $\approx 40 \hmpc$.}
\label{fig:ratio_errors_Pois_FtF_Jack_xis}
\end{figure}

The ratio of Poisson to jackknife errors, Poisson to `field-to-field'
errors, and the `field-to-field' to jackknife errors are given in
Figure~\ref{fig:ratio_errors_Pois_FtF_Jack_xis}. As can be seen, all
error estimators are comparable on scales $ \lesssim 10 \hmpc $, while
at larger scales than this the jackknife and `field-to-field' errors are
considerably larger than the simple Poisson estimates.  The magnitude
of the `field-to-field' and jackknife errors are very similar from the
smallest scales considered here, up to  $\approx 40 \hmpc$.  This
behaviour  has been noted in other correlation function work,
e.g. \citet{daAngela05}. We also note that field-to-field and jackknife
errors are more comparable in size, regardless of scale.  Hence,
the errors that are quoted on all correlation functions 
from here on are the square roots of the variances from the jackknife
method, {\it except} for the case of the angular correlation function,
$w(\theta)$, where we quote the ``Field-to-field'' error. 

\begin{table}
\begin{center}
\setlength{\tabcolsep}{4pt}
\begin{tabular}{ccrrr}
\hline
\hline
Area Name & RA(J2000) range/$^{\circ}$ & LRGs & Randoms 
                                             & $\rho_{rd}/\rho_{LRG}$\\
\hline
 a        & 123.0 - 144.0              &   617 &  10 745 & 17.41  \\ 
 b        & 150.0 - 168.0              & 1 837 &  35 449 & 19.30  \\ 
 c        & 185.0 - 193.0              &   572 &  14 484 & 25.32  \\ 
 d        & 197.0 - 214.0              & 1 723 &  34 373 & 19.95  \\ 
 e        & 218.0 - 230.0              & 1 246 &  24 849 & 19.94  \\
 s06      & 309.2 - 330.0              &   745 &  12 457 & 16.72  \\
 s25      & 330.0 - 360.0              &   876 &  18 499 & 21.12  \\
 s48      &   0.0 -  30.0              &   658 &  13 516 & 20.54  \\
 s67      &  30.0 -  59.7              &   382 &   8 749 & 22.90  \\ 
\hline
Entire Survey    &                     & 8 656 & 173 120 & 20.00  \\
\hline
\hline
\end{tabular}
\caption[The 2SLAQ LRG Survey; Names and Right Ascension 
ranges for the $N=9$ sections used when calculating 
the field-to-field errors.]
{The 2SLAQ LRG Survey; Names and Right Ascension 
ranges for the $N=9$ sections used when calculating 
the field-to-field errors.}
\label{tab:The_Fields_for_errors}
\end{center}
\end{table}

\subsection{Measuring $\xisp$}
Having described how we calculate galaxy-galaxy separations
in redshift-space in order to measure $\xis$, we can now study
the clustering perpendicular, $\sigma$, and parallel, $\pi$, to the 
line of sight. 
We work out the co-moving distance, $r_{c}$, to our object, which
is equal to the distance parallel to the line of sight i.e. a $\pi$
value. Thus, already knowing the redshift-space separation, {\it{s}}, 
we can use 
\begin{equation}
s^2 = \sigma^2 + \pi^2 
\end{equation}
to find $\sigma$. At this point it should be noted that $\sigma$ 
is sometimes designated by $r_{p}$, where $r_{p} \equiv \sigma$. For this 
paper we shall continue to use $\sigma$ for the perpendicular separation.
Closely following \citet{Hoyle02}, $\xisp$ can be estimated in a 
similar way to $\xis$. A catalogue of points, that have the same radial 
selection function and angular mask as the data but are {\it{unclustered}},
is used to estimate the effective volume of each bin. As stated above, 
the unclustered, random catalogue also contains 20 times more points than 
the data. 
The $DD(\sigma,\pi)$, $DR(\sigma,\pi)$ and the $RR(\sigma,\pi)$, 
where again $D$ stands for data LRG and $R$ stands for random, 
counts in each $\sigma$ and $\pi$ bins are found and the 
Landy-Szalay estimator 
\begin{eqnarray}
 \xi_{LS}(\sigma,\pi) =  \frac{\langle DD(\sigma,\pi)  \rangle - 
                               \langle 2DR(\sigma,\pi) \rangle +
			       \langle RR(\sigma,\pi) \rangle}
                              {\langle RR(\sigma,\pi) \rangle},
\end{eqnarray}
is used to find $\xisp$, with bins of 
$\delta\log(\sigma / \hmpc) = \delta\log(\pi / \hmpc) = 0.2$.
Again, we compute three types of errors to use as a guide; 
Poisson, ``Field-to-field'' and Jackknife errors are calculated for
$\xisp$ as in equations \ref{equ:xierr_Poisson} to \ref{equ:xierr_jack}.
Again, after comparing the different $\xisp$ error estimators 
we find that on the scales we are considering, the jackknife error is 
sufficient for our purposes.

\subsection{The Projected Correlation Function, $w_{p}(\sigma)$}

Although we are now in a position to calculate the redshift-space
correlation function, the real-space correlation function, 
$\xir$, which measures the physical clustering of galaxies and is independent 
of redshift-space distortions, remains unknown. 
However, due to the fact that redshift distortion effects only appear
in the radial component, by integrating along the $\pi$ direction, 
we can calculate the projected correlation function, 
\begin{eqnarray}
 w_{p}(\sigma) \;\; = 2 \int_{0}^{\infty}    \xi(\sigma,\pi) d\pi 
\label{eq:Xisigma}
\end{eqnarray}
In practice we set the upper limit on the integral to be 
$\pi\max=70\Mpc$ as at this large-scale, the effect of clustering 
is negligible, while linear theory should also apply.  The effect of
$z$-space distortions due to small-scale peculiar velocities or
redshift errors is also minimal on this scale. 
Changing the value of $\pi\max$ from $25\Mpc$ to $100\Mpc$ makes 
negligible difference in the result.

Due to $w_{p}(\sigma)$ now describing the real-space clustering, the 
integral in Equation \ref{eq:Xisigma} can be re-written in terms
of $\xir$, \citep{Davis83}
\begin{equation}
w_{p}(\sigma) = 2 \int_{\sigma}^{\pi\max} \frac{r\,\xir}
                           {\sqrt{(r^2\,-\,\sigma^2)}}dr .
\label{eq:Xisigma2}
\end{equation}
If we then assume that $\xir$ is a power-law of the form, 
$\xir\,=\,(r/r_0)^{-\gamma}$, equation \ref{eq:Xisigma2} can 
be integrated analytically such that
\begin{equation}
 \frac{w_{p}(\sigma)}{\sigma} = \left(\frac{r_0}{\sigma}\right)^{\gamma} \left[
    \frac {
           \Gamma(\frac{1}{2}) \,  \Gamma(\frac{\gamma-1}{2})
	   }{
           \Gamma(\frac{\gamma}{2})
	   }
\right]
= \left( \frac{r_{0}}{\sigma} \right)^{\gamma} A(\gamma)
\label{eq:wp_div_sigma}
\end{equation}
where $A(\gamma)$ represents the quantity inside the square brackets and
$\Gamma(x)$ is the Gamma function calculated at $x$.
We now have a method for calculating the real-space correlation
length and power-law slope, denoted $r_0$ and $\gamma$ respectively.

\subsection{The Real-space Correlation Function, $\xir$}

Using the projected correlation function, $w_{p}(\sigma)$, it
is now possible to find the $r_0$ and $\gamma$ for the 
real-space correlation function. 
However, if one does not assume a power-law $\xir$, it is still
possible to estimate $\xir$ by directly inverting $w_{p}(\sigma)$. 
Following \citet{Saunders92} we can write
\begin{equation}
  \xir = -\frac{1}{\pi}\int^{\infty}_r 
         \frac{(dw(\sigma)/d\sigma)}
         {(\sigma^2 - r^2)^{\frac{1}{2}}}  d\sigma .
\end{equation}
Assuming a step function for $w_{p}(\sigma)=w_i$ in bins centred on
$\sigma_i$, and interpolating between values,
\begin{equation}
  \xi(\sigma_i) = -\frac{1}{\pi}\sum_{j\geq i}
  \frac{w_{j+1}-w_j} {\sigma_{j+1}-\sigma_j}
  \ln \left( \frac{\sigma_{j+1}+\sqrt{\sigma^2_{j+1} - \sigma^2_i}}
	          {\sigma_j + \sqrt{\sigma^2_j - \sigma^2_i}}\right)
\label{eq:xir_interp}
\end{equation}
for $r = \sigma_i$. 
We shall be utilising this interpolation method to check whether a power-law
description is valid for our 2SLAQ Survey data and, if so, what values
the parameters $r_{0}$ and $\gamma$ take.

%%%%%%%%%%%%%%%%%%%%%%%%%%%%%%%%%%%%%%%%%%%%%%%%%%%%%%%%%%%%%%%%%%%%
%SECTION 3  SECTION 3  SECTION 3  SECTION 3  SECTION 3  SECTION 3  %
%SECTION 3  SECTION 3  SECTION 3  SECTION 3  SECTION 3  SECTION 3  %
%SECTION 3  SECTION 3  SECTION 3  SECTION 3  SECTION 3  SECTION 3  %
%%%%%%%%%%%%%%%%%%%%%%%%%%%%%%%%%%%%%%%%%%%%%%%%%%%%%%%%%%%%%%%%%%%%
\section{Results}

\subsection{The LRG Angular Correlation Function, $w(\theta)$}
\label{sec:w_theta}

\begin{figure}
  \centering
  \centerline{\psfig{file=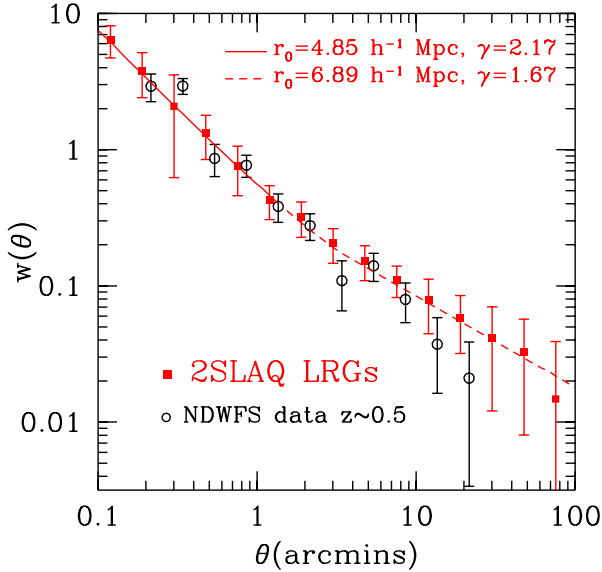,width=8cm}}
  \caption
   [The angular correlation function, $w(\theta)$ from the 2SLAQ LRG Input
   Catalogue]
   {The angular correlation function, $w(\theta)$ from the 2SLAQ
    input catalogue containing \hbox{25 795} LRG targets (solid, red squares). 
    Clear evidence is seen for a change of power-law slope on $\sim 2$arcmin 
    scales which is equivalent to $\approx 1 \hmpc$.
    The open (black) circles show the results from the NDWFS at
    $z\sim0.5$ \citep{White07}.}
    \label{fig:w_theta_2SLAQ_labels}
\end{figure}

We first analyse the form of the angular correlation function,
$w(\theta)$.  The full input catalogue contains approximately \hbox{75
000} LRGs mainly from areas in the two equatorial stripes; about 40\%
of this area was observed spectroscopically. As stated in Section 2,
approximately a third of the objects in  the total input catalogue pass
the Sample 8 selection criteria. As well as providing estimates of
fibre collision and other angular incompletenesses, the angular
function is of interest in itself, particularly given the narrow
redshift range from which the sample is derived. We use \hbox{25 795}
``Sample 8'' LRG targets to estimate the $w(\theta)$.  We first note
that the function gives clear indication of a change of slope at
$\theta=2$ arcmin or $\approx 1 \hmpc$ in the $\Lambda$
cosmology. Considering the form of $w(\theta) = A \theta^{1-\gamma}$,
at $\theta<2$ arcmin the slope is $\gamma=-2.17\pm{0.07}$ and at larger
scales the slope is $\gamma=-1.67\pm{0.07}$. Using Limber's formula
from \citet{Phillipps78} and assuming a double power-law form where the
slope changed between -2.17 and -1.67 at $\sim 1.5 \hmpc$ (comoving),
we found in the $\Lambda$ case, a value of $r_0=4.85\pm0.3 \hmpc $ at
small scales and $r_0=6.89\pm0.6\hmpc$ at large scales (see
Fig.~\ref{fig:w_theta_2SLAQ_labels}). We shall check models of this
form against the deprojected correlation function $\xir$ (see
Figure~\ref{fig:xir_2SLAQ_Cov_EdS_div_plfit} below). We find that the
form of this double power-law gives reasonable fits to the data in the
LRG redshift survey, although the large scale slope derived from the
input catalogue $w(\theta)$ appears slightly flatter than in the
semi-projected  and 3-D correlation functions (see below). The reason
for this is not clear, although it could be that $w(\theta)$ is more
sensitive to any artificial gradient in the LRG data.  Thus, we checked
for an angular systematic in the data by calculating the angular
correlation between spectroscopic LRGs that are not at the same
redshift. We find this is consistent with zero and so such systematics
do not explain the flatter slope  for $w(\theta)$ at large-scales. The
most likely explanation is the different fitting ranges for $w(\theta)$
and the semi-projected correlation function.  This test also suggests
that the upturn at $\theta < 2$ arcmins  is a real feature. It will be
seen that $w(\theta)$ gives the strongest evidence of all the
correlation function statistics for non-power-law behaviour in
$\xir$. A similar feature is seen by Zehavi et al in the SDSS MAIN
galaxy sample and to a lesser extent in the SDSS LRG survey. Reports of
such features in galaxy correlation functions go back to
\citet{Shanks83b}. We simply report the existence of this feature in
the LRG data and leave further interpretation for a future
paper. Possible interpretations could include models of halo occupation
distributions (HOD) in the standard model case or the possibility that
it might represent a real feature in the mass distribution in the case
of other models.  We also show results from  \citet[][open, black
circles, Figure~\ref{fig:w_theta_2SLAQ_labels}]{White07} who report on
the angular correlation function as a route to estimating  merger rates
of massive red galaxies. As can be seen, these measurements from the
NOAO Deep Wide-Field Survey \citep[NDWFS;][]{JD99} agree very well with
the 2SLAQ LRG results, though as we shall discuss later,  care always
has to be taken when comparing measurements from galaxy surveys with
different selections.

\subsection{The LRG Redshift-Space Correlation Function, $\xi(s)$}
\begin{figure}
\centering
\centerline{\psfig{file=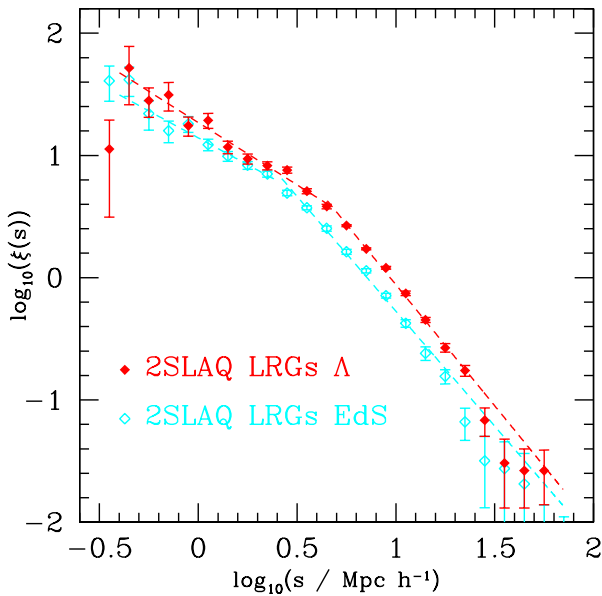,width=8cm}}
\caption{The redshift-space 2-point correlation function, $\xi(s)$ for
the 2SLAQ LRG Survey in a $\Lambda$ cosmology (filled, red diamonds) 
and an Einstein-de Sitter, $\Omm = 1$, cosmology (open, cyan diamonds).
The dashed lines shown are the double power-law best-fit models to 
data with the associated values of $s_{0}$ and $\gamma$ given in 
Table~\ref{tab:plfit2_xis}.}
\label{fig:xis_2SLAQ_Cov_plfit2_EdS}
\end{figure}

\begin{figure}
\centering
\centerline{\psfig{file=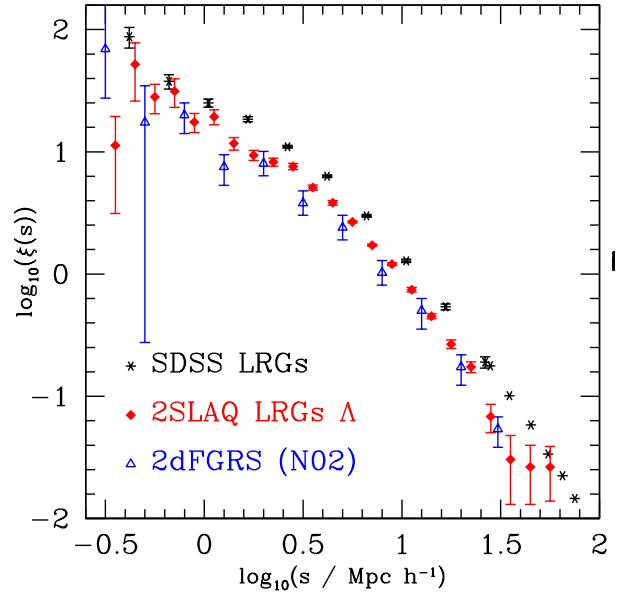,width=8cm}}
\caption{The redshift-space correlation function, $\xi(s)$ for
the 2SLAQ LRG Survey (filled, red, diamonds). 
For comparison, data from the SDSS LRG Survey 
\citep[black stars][]{Zehavi05a, Eisenstein05} and 
the high luminosity early-type 2dFGRS, 
\citep[open blue triangles]{Norberg02a} 
are also plotted.}
\label{fig:xis_2SLAQ_Cov_GRS_Zehavi_Eisenstein}
\end{figure}

Using the above corrections including that for fibre collisions, the
2SLAQ LRG redshift-space 2PCF, $\xis$, is shown in
Figure~\ref{fig:xis_2SLAQ_Cov_plfit2_EdS}. There is clear evidence for
downturns at small scales $\lsim 2.5 \hmpc$ and large scales $\gsim 10
\hmpc$ that are not described well by a single power-law. This turn-over
is consistent with the redshift-space distortion effects one would
expect in a $\xi(s)$ correlation function - namely the ``Finger of God"
effect at small scales due to intrinsic velocity dispersions and
large-scale flattening from peculiar motions due to coherent cluster
in-fall. However, we note that real features in the real-space
correlation function, $\xir$, may also be contributing. We have also
estimated the effect of the integral constraint \citep[$IC$,][]{Peebles80} 
at larger scales. 
Using our global (N$+$S) normalisation of the correlation function, 
we assume a total number of \hbox{8 656} galaxies in a total volume of
$4.5 \times 10^{7} \hmpc^{3}$ and $r_{0}=7.45 \hmpc$. 
Integrating with a  $\gamma = 1.8$ power-law to 20$\hmpc$ gives an
$IC= 3.5 \times 10^{-4}$ and to 100$\hmpc$, an $IC=2.4 \times 10^{-3}$. 
Adding such contributions would make negligible contributions to any of our
correlation function fits.

We now attempt to parameterise the $\xi(s)$ data.  The simplest
model traditionally fitted to correlation  function estimates is a
power law of the form
\begin{equation}
\xi(s)=\left(\frac{s}{s_0}\right)^{-\gamma},
\label{eq:xipowerlaw}
\end{equation}
where $s_0$ is the comoving correlation length, in units of $\hmpc$.
However, with the redshift-space distortion effects being so evident, 
we find that a single-power is insufficient to describe the data
and thus switch to a double power-law model
\begin{equation}
  \xi(s)= \left\{ \begin{array}{ll}
                \left(\frac{s}{s_{1}}\right)^{\gamma_{1}} & s \leq s_{\rm{b}}
                                              \;\;\;  \rm{and}\;\; \\
                \left(\frac{s}{s_{2}}\right)^{\gamma_{2}} & s > s_{\rm{b}}
                \end{array}
        \right.
  \label{eq:xipowerlaw2}
\end{equation}
where $s_{\rm b}$ is the scale of the ``break'' from one power-law
description to the other. This $\xis$ model is used later in
Section~\ref{sec:comso_params}. We fit the double power-law
continuously over the range $0.4 < s < 60 \hmpc$.  We fix the
break-scale at 4.5$\hmpc$ for the $\Lambda$ cosmology and at 2.5$\hmpc$
for the EdS cosmology. We perform a $\chi^{2}$-fit, following the
prescription given by \citet[Chap. 15]{Press92}, to find the best-fit
values for $s_{1}, \gamma_{1}, s_{2},$ and $\gamma_{2}$. We plot the
best fit double-power law models in
Figure~\ref{fig:xis_2SLAQ_Cov_plfit2_EdS} and  quote the values of
$s_{1}, \gamma_{1}, s_{2},$ and $\gamma_{2}$, in
Table~\ref{tab:plfit2_xis}. The errors quoted in Table 3 are only
indicative because no account has been taken of the non-independence of
the correlation function points in deriving the $\xi(s)$ fits.

\begin{table}
\centering
\caption{Values of the redshift-space correlation length and slope for the 
2SLAQ LRG Survey from $\xi(s)$. When a $\Lambda$ cosmology was assumed, 
$s_{b}$ was set at 4.5$\hmpc$. When a EdS cosmology was assumed, 
$s_{b}$ was set at 2.5 $\hmpc$.}
\begin{tabular}{||l|c|c||} \hline
\hline
$\Lambda$                  & $s_{0} < 4.5 \hmpc$   & $s_{0} > 4.5 \hmpc$   \\
\hline
${\rm s}_{0} / \hmpc$      & $17.3^{+2.5}_{-2.0} $ & $ 9.40 \pm 0.19     $ \\
$\gamma$                   & $1.03\pm0.07        $ & $ 2.02 \pm 0.07     $ \\
$\chi^{2}_{min}$ (reduced) & 1.95                  & 1.88                  \\
%d.o.f.                    & 9                     & 10                    \\  
\hline
EdS                        & $s_{0} < 2.5 \hmpc$   & $s_{0} > 2.5 \hmpc$   \\
\hline  
${\rm s}_{0} / \hmpc$      & $20.3^{+9.4}_{-5.0} $ & $ 7.15\pm0.13       $ \\
$\gamma$                   & $0.88\pm0.11        $ & $ 1.88^{+0.05}_{-0.04}$ \\
$\chi^{2}_{min}$ (reduced) & 0.91                  & 3.43                  \\
%d.o.f.                    & 6                     & 12                    \\  
\hline
\hline
\end{tabular}
\label{tab:plfit2_xis}
\end{table}

For comparison, in Figure~\ref{fig:xis_2SLAQ_Cov_GRS_Zehavi_Eisenstein}
results from the SDSS LRG study are reported
\citep{Zehavi05a,Eisenstein05} as well as selected measurements from
the 2dFGRS \citep{Norberg02a}. The 2dFGRS is a blue, $b_{J}$ selected
survey of  generally $\sim L^{*}$ galaxies. However, in
\citet{Norberg02a}, the sample is segregated by luminosity and spectral
type, the latter governed by the $\eta$ parameter \citep{Madgwick03}.
Assuming a conversion of $M^{0.2}_{r} - M_{\bj} \simeq -1.1$,  we
calculate that the faintest 2SLAQ LRGs in our sample have an  $M_{\bj}
\approx -20.5$. Weighting according to number, we thus use the
\citet{Norberg02a}  $-21.00 > M_{\bj} - 5 \log h > -22.00$  and $-20.50
> M_{\bj} - 5 \log h > -21.50$ luminosity ranges from their
``early-type'' volume-limited sample. This is shown by the (blue) open
triangles in Figure~\ref{fig:xis_2SLAQ_Cov_GRS_Zehavi_Eisenstein}.

The 2SLAQ LRG measurement is lower than the SDSS LRG result.  It should
not be concluded that this is evidence of evolution because  although
the SDSS survey is at a lower mean redshift,  it was designed in order
to target generally redder, more luminous LRGs
\citep{Eisenstein01}. The 2SLAQ LRG colour selection criteria is
relatively relaxed for an ``LRG'' survey, leading to bluer and less
luminous galaxies making it into our sample.  We note here that it is
non-trival comparing clustering amplitudes and bias strengths for
surveys with (sometimes very) different colour/magnitude/redshift
selections. As such, a more detailed analysis of the clustering
evolution for SDSS and 2SLAQ LRGs is presented in Wake et al. (2007, in
prep.).

The 2dFGRS $M_{\bj} < -20.5$, early-type sample is at least
approximately matched in terms of luminosity to the 2SLAQ LRGs. Once we
have determined the linear bias parameter $b$ for the $z = 0.55$ 2SLAQ
LRGs, we shall be able to use a simple model of bias evolution, to
compare these low redshift 2dFGRS and 2SLAQ LRG results.

\subsection{The Projected Correlation function, $w_p(\sigma)$}
\begin{figure}
\centering
\centerline{\psfig{file=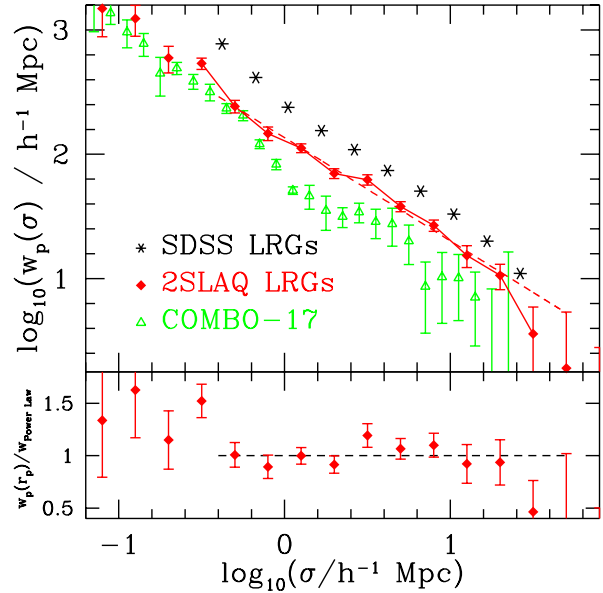,width=8cm}}
\caption{The 2SLAQ LRG projected correlation function, $w_{p}(\sigma)$, with 
errorbars  from the ``Jackknife'' estimates (solid, red diamonds).
The dashed line is the power-law that gives the best fitting line 
from the $\chi^{2}$ analysis (see Table~\ref{tab:plfit1_wp}). 
The measurements from the SDSS LRGs \citep{Zehavi05a} are shown as a guide, 
with the SDSS errors being of comparable size to the plotted stars.
The open (green) triangles are from COMBO-17 Red Sequence 
\citep{Phleps06}.
The lower panel shows the 2SLAQ LRG
$w_p(\sigma)$ measurements divided by this best-fitting power law 
with the dashed line covering $0.4 < \sigma < 50 \hmpc$.}
\label{fig:wp_div_plfit_2SLAQ_Cov}
\end{figure}

Again, after applying coverage, spectroscopic and fibre collision
corrections, the projected correlation function, $w_p(\sigma)$, is
presented in Figure~\ref{fig:wp_div_plfit_2SLAQ_Cov}. We again fit a
single power-law to the 2SLAQ data and find that for the $\Lambda$
cosmology, a single power-law is an adequate description, returning a
reduced $\chi^{2}$ = 1.17 over $0.4 < \sigma < 50 \hmpc$. Over the
wider range of $0.1 < \sigma < 50 \hmpc$, the $\chi^{2}$ increases to
1.71. Thus the projected correlation function appears to deviate from a
single power law at small scales in the way described in
Section~\ref{sec:w_theta}. The results for $r_0$ and $\gamma$ assuming
a single power-law are given in Table~\ref{tab:plfit1_wp}. The errors
are taken from jack-knife estimates found by dividing the survey into
32 subareas.

\begin{table}
\centering
\caption{Values of the projected correlation function, $w_{p}(\sigma)$,
correlation length and slope for the 2SLAQ LRG Survey.
In the $\Lambda$ model, fits were performed over the range 
$0.4 < \sigma < 50.0 \hmpc$, whereas for the EdS model, fits were performed 
over $0.25 < \sigma < 40.0 \hmpc$.
The value of $r_0$ was found using equation~\ref{eq:wp_div_sigma}.}
\begin{tabular}{||l|c|c||} \hline
\hline
                           & $\Lambda$          &   EdS               \\
\hline
$r_0 / \hmpc$              & $7.30\pm0.34 $     &   $5.40\pm0.31$     \\
%$\gamma$                   & $1.84\pm0.049$    &   $1.77\pm0.051$    \\
$\gamma$                   & $1.83\pm0.05$      &   $1.82\pm0.06$     \\
$\chi^{2}_{min}$ (reduced) & 1.17               &   1.39              \\
%d.o.f.                    & 9                  &   9                 \\       
\hline
\hline
\end{tabular}
\label{tab:plfit1_wp}
\end{table}

This power-law deviation in the projected correlation function is in
line with recent results seen in other galaxy surveys, e.g. the SDSS
MAIN sample (\citet{Zehavi04}, not plotted) and the SDSS LRGs
\citep{Zehavi05a}. A ``shoulder'' is reported in these studies around
$\sim 1 \hmpc$ scales. This feature is currently believed to be a
consequence of the transition from the measuring of galaxies that
reside within the {\it same} halo (the ``one-halo'' term) to the
measuring of galaxies in {\it separate} haloes (the ``two-halo'' term).
Dips in the projected correlation function are a  major prediction of
HOD models. Thus for the 2SLAQ LRG Survey, we set a fiducial model,
based on our best-fitting single power-law model of $w_{p}(\sigma)$ and
find that if we divide the data out by this model, the results (bottom
panel, Figure \ref{fig:wp_div_plfit_2SLAQ_Cov}) are potentially
comparable to the \citet{Zehavi05a} results (their Figure 11).  Despite
the fact that our LRG sample is at higher redshifts and extends to
lower luminosities, the form of the projected correlation function
appears close to that seen in the SDSS LRG sample, although at lower
amplitude.  We conclude that the 2SLAQ LRG correlation function changes
slope in  similar fashion to the SDSS LRG semi-projected correlation
function.

Continuing with $w_{p}(\sigma)$, we compare the 2SLAQ LRGs with the
COMBO-17 Survey. COMBO-17 (Classifying Objects by Medium-Band
Observations, \citet{Wolf01} uses a combination of 17 filters to obtain
photometric redshifts accurate to $\sigma_{z} /(1+z) \simeq 0.01$ for
the brightest ($R_{\rm Vega} < 20$ mag) objects. This is a comparable
sample to our own in that it covers the same redshift range ($0.4 < z <
0.8$), but care must be taken when comparing the results; although the
COMBO-17 galaxies described here are defined as Red Sequence, on the
whole they  will not be LRGs and will have a fainter magnitude and
different colour  selection.  Figure~\ref{fig:wp_div_plfit_2SLAQ_Cov}
gives the projected correlation function of the 2SLAQ LRGs and red
COMBO-17 galaxies from  \citet{Phleps06} (assuming a flat $\Lambda$
cosmology). The change in slope is clearly seen in COMBO-17 and indeed
is modelled successfully with a HOD prescription \citep{Phleps06}.  The
upturn in slope in COMBO-17 versus 2SLAQ seems to occur on slightly
different scales ($\simeq 1-2 \hmpc$ versus $\simeq 5 \hmpc$) and is
more dramatic than for either of the LRG samples. The errors on the
COMBO-17 data are also much greater. Whether the differences are real,
caused by the fainter magnitude of the COMBO-17 galaxies, or whether
they are due to anomalies caused by the photometric redshifts, remains
unclear.

\subsection{The Real-Space Correlation function, $\xi(r)$}
\begin{figure}
\centering
\centerline{\psfig{file=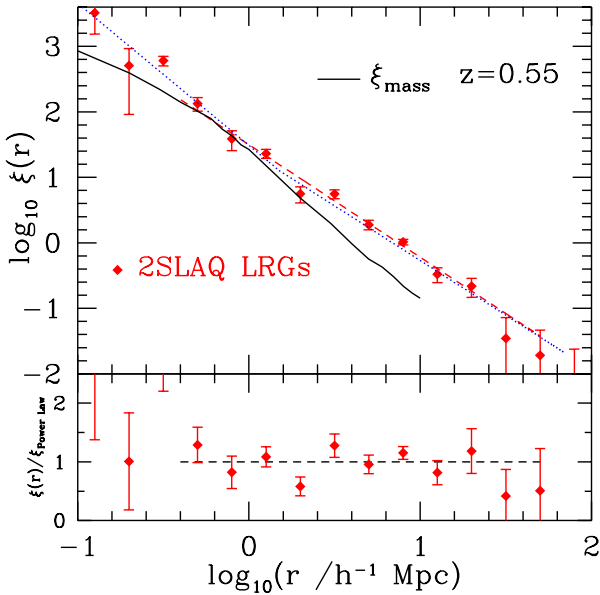,width=8cm}}
\caption{The real-space 2-point correlation function for the 2SLAQ LRG Survey 
(filled, red, diamonds) for the $\Lambda$ cosmology.
The best-fit single power-law with 
$r_{0} = 7.45\pm0.35$ and $\gamma = 1.72\pm{0.06}$ is
given by the dashed (red) line. 
The double power-law fit reported for the angular correlation, $w(\theta)$, 
in Section 3.1, is shown by the dotted (blue) line.
The solid (black) line is a theoretical prediction for the 
$\xi_{\rm mass}(z=0.55)$ using the simulations from \citet{Colin99}. 
These models have $(\Omm$,$\Omlam)=$ $(0.3,0.7)$, $h=0.7$ and a 
$\sigma_{8} = 1.0$.
We shall return to this in Section 4. 
The lower panel shows the 2SLAQ LRG $\xi(r)$ measurements 
(assuming a $\Lambda$ cosmology) divided by this best-fitting power 
law with the dashed line covering $0.4 < \sigma < 50 \hmpc$.}
\label{fig:xir_2SLAQ_Cov_EdS_div_plfit}
\end{figure}

Having reported the clustering of 2SLAQ LRGs using the $z$-space correlation 
function, $\xi(s)$ and the projected correlation function, $w_{p}(\sigma)$ we 
now use the methods quoted in Section 2 to estimate the real-space correlation 
function, $\xir$. We show this in
Figure~\ref{fig:xir_2SLAQ_Cov_EdS_div_plfit}. 

\begin{figure}
\centering
\centerline{\psfig{file=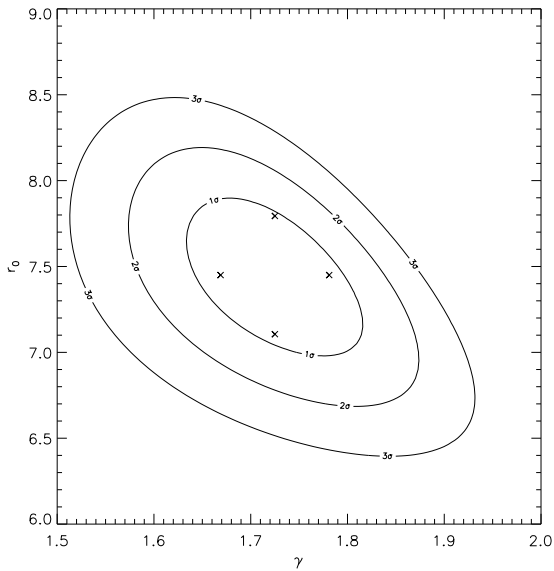,width=8cm}}
\caption{The joint 2 parameter fits on $r_{0}$ and $\gamma$ for $\xi(r)$.
The contours show the $\delta \chi^{2}=(2.3, 6.17, 11.8)$ corresponding to 
1, 2 and 3$\sigma$. 
The crosses show the deviations in $r_{0}$ and $\gamma$ that 
we find from the 32 best-fitting single power-law using the jackknife samples.}
\label{fig:fit_s0_gamma_full_rv1_xir_LS_jacks_paper}
\end{figure}

\begin{table}
\centering
\caption{Values of the correlation length and slope for the 
2SLAQ LRG Survey from the real-space correlation function, $\xir$. 
Model fits were performed over the range $0.4 < r < 50 \hmpc$ for
the $\Lambda$ cosmology and over the range $0.25 < r < 40 \hmpc$ for
the EdS cosmology.}
\begin{tabular}{||l|c|c||} \hline
\hline
                           & $\Lambda$           &    EdS \\
\hline
$r_{0} / \hmpc$            & $7.45\pm0.35$       &  $ 5.65\pm0.41$ \\
$\gamma$                   & $1.72\pm0.06$       &  $ 1.67\pm0.09$ \\
$\chi^{2}_{min}$ (reduced) & 1.73                &  0.62           \\
%d.o.f.                    & 9                   &  9              \\        
\hline
\hline
\end{tabular}
\label{tab:fitpowpar}
\end{table}
Again, we attempt to fit simple power-law models to our $\xir$ data in order
to find values for the real-space correlation length and slope, $r_0$ and
$\gamma$, respectively. For $\xi(r)$ we attempt to take into account the 
information presented in the covariance matrix by estimating $\chi^{2}$ fits to
model $\xi(r)$ values such that
\begin{equation}
\chi^2 = \sum_{i,j}[\bar{\xi}(r_{i}) - \xi_{m}(r_{i})] \,
                   C_{i,j}^{-1} \,
                   [\bar{\xi}(r_{j}) - \xi_{m}(r_{j})]
\label{eq:Chi}
\end{equation}
where $C_{ij}^{-1}$ is the inverse matrix of the covariance matrix and the 
subscripts $i$ and $j$ are indicies of separation bins. However, as has been 
reported in previous clustering analyses (e.g. \citet{Zehavi02, Scranton02}), 
the calculated covariance matrix is rather noisy with 
anti-correlations between points (contary to theoretical expectations). 
Therefore, when calculating the best-fitting models, we perform a simple 
$\chi^{2}$ fit as before, without the covariances, and take only the 
variances into account. As before, we fit over the scales 
$0.4 \leq r \leq 50.0 \hmpc$.
For the case of the real-space correlation function, we again find that
a single power-law may not fit the data well with the best-fit values
(and related reduced $\chi^{2}$) given in Table~\ref{tab:fitpowpar}. We
find a value of $\gamma$ to be  $1.72\pm0.06$ and a
correlation length of $r_{0} = 7.45\pm0.35$ (assuming a $\Lambda$ cosmology). 
The errors on these parameters are estimated from considering the 
1$\sigma$ deivation from the minimized $\chi^{2}$ on the 1-parameter fits. 
However, care has to be taken when quoting the best fit values for the 
joint 2-parameter fits which are shown in
Figure~\ref{fig:fit_s0_gamma_full_rv1_xir_LS_jacks_paper}. 
Here we find the values of $\delta \chi^{2}$ which correspond to the 
1, 2 and 3$\sigma$ levels for a 2-parameter fit. Also shown in
Fig.~\ref{fig:fit_s0_gamma_full_rv1_xir_LS_jacks_paper} are the values
for the deviations in $r_{0}$ and $\gamma$, if we find the 32
best-fitting single power-law parameters from the jackknife samples.
Jackknife appears to confirm the $\chi^{2}$ error analysis with the assumption 
of Gaussian errors in Fig.~\ref{fig:fit_s0_gamma_full_rv1_xir_LS_jacks_paper}. 
This is somewhat surprising since we have
ignored the covariance between correlation function points in creating 
Fig.~\ref{fig:fit_s0_gamma_full_rv1_xir_LS_jacks_paper}. 
The explanation may be that the fit at the minimum is still poor due to
the deviant point at $2 \hmpc$ in Fig.~\ref{fig:xir_2SLAQ_Cov_EdS_div_plfit} 
and this causes the error contours in 
Fig.~\ref{fig:fit_s0_gamma_full_rv1_xir_LS_jacks_paper} 
to be larger than they would be in the absence of the deviant
point. Including the full covariance matrix, the $\Delta \chi^{2}$ produces
error contours significantly smaller than those in 
Fig.~\ref{fig:fit_s0_gamma_full_rv1_xir_LS_jacks_paper} and also the
jackknife errors, even though the $\chi^{2}$ at minimum remained the same.
Overall we take the errors in 
Fig.~\ref{fig:fit_s0_gamma_full_rv1_xir_LS_jacks_paper} supported by the 
jackknife estimates as being reasonably representative of the real error.

Now armed with our best-fitting single power-law model for $\xi(r)$, and
we can proceed and see if modelling the redshift-space distortions introduced 
into the clustering pattern reveals anything about cosmological parameters.

%%%%%%%%%%%%%%%%%%%%%%%%%%%%%%%%%%%%%%%%%%%%%%%%%%%%%%%%%%%%%%%%%%%%
%SECTION 4  SECTION 4  SECTION 4  SECTION 4  SECTION 4  SECTION 4  %
%SECTION 4  SECTION 4  SECTION 4  SECTION 4  SECTION 4  SECTION 4  %
%SECTION 4  SECTION 4  SECTION 4  SECTION 4  SECTION 4  SECTION 4  %
%%%%%%%%%%%%%%%%%%%%%%%%%%%%%%%%%%%%%%%%%%%%%%%%%%%%%%%%%%%%%%%%%%%%
\section{LRG clustering and Cosmological Implications}
\label{sec:comso_params}

Having calculated the $z$-space, projected and real-space correlation 
functions for the 2SLAQ Luminous Red Galaxies, we can now turn our attention 
to using these results to see if we can determine cosmological parameters. 

\subsection{The $\xi(\sigma, \pi)$ LRG Measurements}
\begin{figure}
\centering
\centerline{\psfig{file=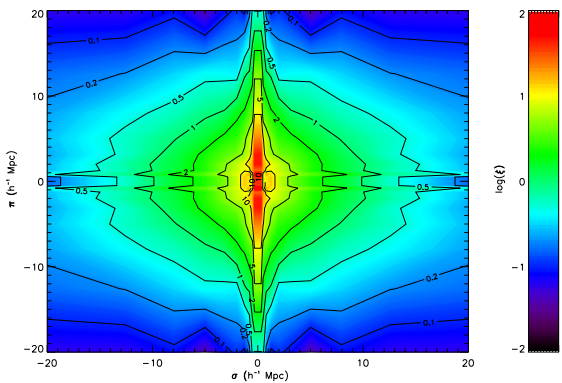,width=8cm}}
\caption{The $\xisp$ contour plot for the 2SLAQ LRG Survey, assuming
a $\Lambda$ cosmology of $(\Omm,\Omlam)=(0.3,0.7)$.
The ``Finger-of-God'' effects, i.e. elongation of 
contours in the $\pi$ direction at small ($\lsim 1 \hmpc$) scales, are
clearly seen. (The spikes at small $\pi$ are a plotting artefacts).}
\label{fig:xi_si_pi_plot_full}
\end{figure}

\begin{figure}
\centering
\centerline{\psfig{file=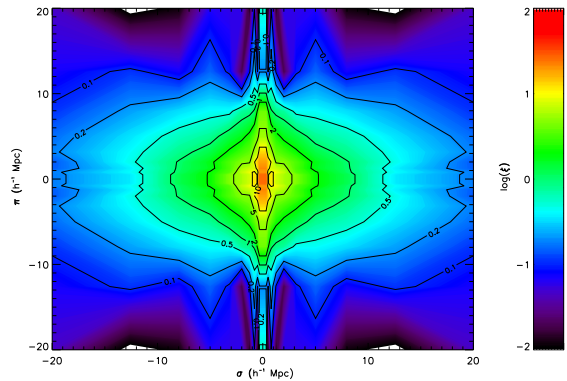,width=8cm}}
\caption{The $\xisp$ contour plot for the 2SLAQ LRG Survey, with
a $\Omega_{\rm m}=1.0$, EdS cosmology.}
\label{fig:xi_si_pi_plot_LS_EdS}
\end{figure}

Results for the 2-D clustering of 2SLAQ LRGs are shown in the
$\xisp$ plots of Figures~\ref{fig:xi_si_pi_plot_full}
and~\ref{fig:xi_si_pi_plot_LS_EdS}.

Galaxy peculiar velocities lead to distortions in the $\xisp$ shape. 
The predominant effect at large scales in $\sigma$ is the coherent infall 
that causes a flattening of the $\xisp$ contours along the parallel
$\pi$ direction and some elongation along the perpendicular $\sigma$ 
direction. At small $\sigma$, the random peculiar motions of the galaxies
cause an elongation of the clustering signal along the $\pi$ direction - the
so-called ``Fingers-of-God'' effect. 
From the measurements of these effects, determination of the coherent infall 
into clusters, given by the parameter $\beta$, and the 
pairwise velocity dispersion, $\wrms$, can be made. This calculation shall 
be performed in Section~\ref{sec:beta_and_pairwise}.
Geometric distortions also occur if the cosmology assumed to convert the 
observed galaxy redshifts is not the same as the true, underlying cosmology 
of the Universe. The reason for this is because the cosmology dependence of 
the separations along the redshift direction is not the same as for the 
separations measured in the perpendicular direction \citep{AP79}.
We note that modelling the geometric distortions and comparing to 
the presented data can yield information on cosmological parameters.

We shall closely follow the methods of Hoyle et al. (2002) and 
da \^{A}ngela (2005), hereafter H02 and dA05, respectively. 
In this section, we first discuss large-scale, 
linear and small-scale non-linear $z$-space distortions and how they are
parameterised by $\beta$ and $\wrms$ respectively. 
We then use $\beta$ to find the bias of LRGs at the survey redshift. 
Next, we employ information gained in studying the geometric distortions 
to perform the ``Alcock-Paczynski Test'' as one route to calculating 
cosmological parameters. 
However, we realise there is a degeneracy in the $(\beta, \Omm)$ plane 
with this approach and thus employ further constraints from the 
evolution of LRG clustering to break this degeneracy.

\subsection{Redshift-space distortions, $\beta$ and 
pairwise velocities}\label{sec:beta_and_pairwise}
When measuring a galaxy redshift, one is actually measuring a sum 
of velocites.\footnote{This section strongly follows \citet{Hawkins03} and
\citet{Croom05}.}  
The total velocity comes from the Hubble expansion plus the motion induced
by the galaxy's local potential, where this second term is coined
the ``peculiar velocity'', i.e. 
\begin{equation}
v_{\rm Tot} = v_{H} + v_{\rm pec}
\label{eq:v_tot}
\end{equation}
The peculiar velocity itself contains two terms, 
\begin{equation}
v_{\rm pec} =  v_{\rm rand} + v_{\rm CI} 
\label{eq:v_pec}
\end{equation}
The first term, $v_{\rm rand} $ is due to the small-scale random motion of 
galaxies within clusters. The second term, $v_{\rm CI}$ is the component due 
to coherent infall around clusters, where the infall is caused by the 
streaming of matter from underdense to overdense regions; this
leads to a ``flattening'' in the perpendicular $\sigma$-direction away 
from equi-distant contours in $\xisp$. This extension is
parameterised by $\beta$, which takes into account the 
large-scale effects of linear $z$-space distortions.
Kaiser (1987) showed that, assuming a pure power-law model for
the real-space correlation function (which is fair for the 2SLAQ LRG data), 
one can estimate $\beta$ in the linear regime using
\begin{equation}
\xi(s)= \xi(r) \left({ 1+\frac{2}{3}\beta + \frac{1}{5}\beta^{2}} \right).
\label{eq:xis_div_xir}
\end{equation}
and more generally
\begin{equation}
\xi(\sigma,\pi) =   \left[  1+\frac{2(1-\gamma\mu^2)}{3-\gamma}\beta \\
                  +  \frac{3-6\gamma\mu^2+\gamma(2+\gamma)\mu^4}{(3-\gamma)
                     (5-\gamma)}\beta^2\right]\xi(r), 
\label{eq:xisigmapilin}
\end{equation} 
where  $\mu$ is the cosine of the angle between $r$ and $\pi$ 
(the distance along the line of sight), 
and $\gamma$ is slope of the power law \citep{Matsubara96}.

Even though the ``Kaiser Limit'' is a widely 
used method for estimating $\beta$, drawbacks using this approach, 
under the assumption of Gaussianity, have been known for some time 
\citep{Hatton98}. Scoccimarro (2004) has recently reported on 
the limitations of assuming a Gaussian distribution in
the pairwise velocity dispersion $\sigma_{12}$, even at very large scales. 
Scoccimarro's argument is that even at large scales, linear theory 
cannot be applied since one still has the effect of galactic motions 
induced on sub-halo scales i.e. galaxies that are separated by very large
distances are still ``humming'' about inside their own dark mattter haloes. 
Thus for the remainder of the paper, we make a note of the new 
formalism in Scoccimarro (2004), but continue to 
use the Kaiser Limit, acknowledging its short-comings. We justify this
by noting that we need better control on our `1st order' statistical and 
systematic errors before applying the `2nd order' Scoccimarro corrections.
Future analysis may use the 2SLAQ LRG and QSO sample to make comparisons
for small and large scale effects in the redshift distortions using both
the new Scoccimarro expression as well as the Kaiser Limit.

The small-scale random motions of the galaxies, $v_{\rm rand}$, 
leads to an extension in the $\pi$-direction of $\xisp$. 
We denote the magnitude of this extension by $\wrms$ 
($\equiv \sigma_{12}$); this is usually expressed in 
a Gaussian form (e.g. dA05) 
\begin{equation}
f(w_{z}) = \frac{1} {\sqrt{2 \pi} \langle w_{z}^{2} \rangle^{1/2}}
           \rm{exp} \left( - \frac{1}{2} \frac{ | w_{z} |  }
                    {\langle w_{z}^{2} \rangle^{1/2}}
                    \right).
\label{eqn:exp_pairwise}
\end{equation}
Now we can combine these small-scale non-linear $z$-space distortions with
the Kaiser formulae, and hence the full model for $\xisp$ is given by
\begin{equation}
\xi(\sigma,\pi) = \int^{\infty}_{-\infty} \xi^{\prime}
                  [\sigma, \pi - w_{z}(1+z) / H(z)] f(w_{z}) dw_{z}
\label{eqn:twenty-seven}
\end{equation}
where $\xi^{\prime}[\sigma, \pi - w_{z}(1+z) / H(z)]$ is given by 
equation~\ref{eq:xisigmapilin} and 
$f(w_{z})$ by equation~\ref{eqn:exp_pairwise}.
Using these expressions and our 2SLAQ LRG data, we can calculate 
$\beta$ and $\wrms$ for the LRGs.   
At this juncture, it is important to note the scales we consider in our 
model. As can be seen from the data presented in Section 3, a power-law fits 
the data best on scales from 1 to 20 $\hmpc$. 
Thus, when computing the full model for $\xisp$ 
(equation~\ref{eqn:twenty-seven}), 
we only use data with $1 < \sigma < 20  \hmpc$ and $1 < \pi < 20 \hmpc$
(as shown in Figures~\ref{fig:xi_si_pi_plot_full} and 
\ref{fig:xi_si_pi_plot_LS_EdS}).

Returning to Kaiser (1987), the value of $\beta$ can be used to determine
the bias, $b$, of the objects in question,
\begin{equation}
\beta \simeq \frac{\, \Omm^{\,0.6}}{b}
\label{eqn:beta_Omegamb}
\end{equation}
provided you know the values of $\Omm$, where  $\Omm(z)$
is given by
\begin{equation}
\Omm(z) = \frac{\Omm^{0}(1+z)^{3}}
                     {\Omm^{0}(1+z)^{3}+\Omega_{\Lambda}^{0}},
\label{equation:omegam_z0.55}
\end{equation}
for a flat universe.
The importance of the bias is that it links the visible 
galaxies to the underlying (dark) matter density fluctuations,
\begin{equation}
\delta_{g} = b \;\; \delta_{m}
\end{equation}
where the $g$ and the $m$ subscripts stand for galaxies and mass
respectively.  However, the precise way in which
galaxies trace the underlying matter distribution is still poorly understood.
Recent work by
e.g. \citet{Blanton06}, \citet{Schulz06}, \citet{Smith07}
and \citet{Coles07} suggests that bias is
potentially scale-dependent and we note that we do not take this into
account in the current analysis. 
Thus, for our purposes, we restrict ourselves to the
very simple relation,  $\xi_{g} = b^{2} \xi_{m}$, where $b$ is the
linear bias term and leave further investigation of the bias for 
massive galaxies  at intermediate redshift to a future paper.
On the above model assumptions we now proceed to 
estimate the cosmological parameters, $\Omm$ and $\beta$.

\subsection{Cosmological Parameters from $\xisp$ models.}
The ratio of observed angular size to radial size 
varies with cosmology. If we have an object which is known to be isotropic, 
i.e. where transverse and radial intrinsic size are the same,  
fixing the ratio of the intrinsic radial and transverse distances yields a 
relation between the measured radial and transverse distances depending 
on cosmological parameters. This comparison is often called the
``Alcock-Paczynski" test (Alcock \& Paczynski 1979; Ballinger, Peacock \&
Heavens 1996). In order to perform this test, we assume 
galaxy clustering is, on average, isotropic and we compare
data and model cosmologies. Following H02 and dA05, 
for the following sections, we define several terms.\\

(i) The Underlying cosmology - this is the true, underlying, 
unknown cosmology of the Universe. 

(ii) The Assumed cosmology - the cosmology used when measuring 
the two-point correlation function and $\xi(\sigma,\pi)$ from the 2SLAQ
LRG survey.\ Initially in a redshift survey, the only information
available is the object's position on the sky and
its redshift. In order to convert this into a physical separation, 
you must assume some cosmology. As was mentioned earlier, 
we have considered two Assumed cosmologies, 
the $\Lambda$ $(\Omega_{\rm m}$,
$\Omega_{\Lambda})$ = (0.3,0.7) and
the EdS $(\Omega_{\rm m},\Omega_{\Lambda})$ = (1.0,0.0) cases. 

(iii) The Test Cosmology - the cosmology used to generate the
model predictions for $\xisp$ which are then translated into the
assumed cosmology. \\ 

We compare the geometric distortions in both 
the data and the model relative to the {\it same Assumed} cosmology.
Thus, the key to this technique lies in the fact that when the Test cosmology
matches the Underlying cosmology, the distortions introduced
into the clustering pattern should be the same in model as in the data. 
The model should then provide a good fit to the data, {\it providing the 
redshift-space distortions have been properly accounted for}.
We can then endeavour to find values of $\Omega_{\rm m}$ and $\beta$.
We assume that for all further discussions, the cosmologies described 
are spatially flat and choose to fit the variable $\Omega_{\rm m}^{0}$, 
hence fixing $\Omega_{\Lambda}^{0} = 1 - \Omega_{\rm m}^{0}$.

The relation between the separations $\sigma$ and $\pi$ in the Test and 
Assumed cosmologies (referred to by the subscripts $t$ and $a$ respectively)
is the following (Ballinger et al. 1996, HO2, dA05):

\begin{equation}
\sigma_{t} = f_{\perp} \sigma_{a} = \frac{B_{t}}{B_{a}} \sigma_{a}
\label{equation:sig_cosmol}
\end{equation}

\begin{equation}
\pi_{t} = f_{\parallel} \pi_{a} = \frac{A_{t}}{A_{a}} \pi_{a}
\label{equation:pi_cosmol}
\end{equation}
where $A$ and $B$ are defined as follows (for spatially flat cosmologies):

\begin{equation}
A = \frac{c}{H_{0}}\frac{1}{\sqrt{\Omega_{\Lambda}^{0}+\Omm^{0}(1+z)^{3}}}
\label{equation:A_cosmol}
\end{equation}

\begin{equation}
B = \frac{c}{H_{0}}\int_{0}^{z}\frac{dz'}{\sqrt{\Omega_{\Lambda}^{0}+\Omm^{0}(1+z')^{3}}}.
\label{equation:B_cosmol}
\end{equation}
In the linear regime, the correlation function in the assumed cosmology 
will be the same as the correlation function in the test cosmology, 
given that the separations are scaled appropriately. i.e.:
\begin{equation}
\xi_{t}(\sigma_{t},\pi_{t}) = \xi_{a}(\sigma_{a},\pi_{a}).
\label{equation:xisp_cosmol}
\end{equation}

\begin{figure}
\centering
\centerline{\psfig{file=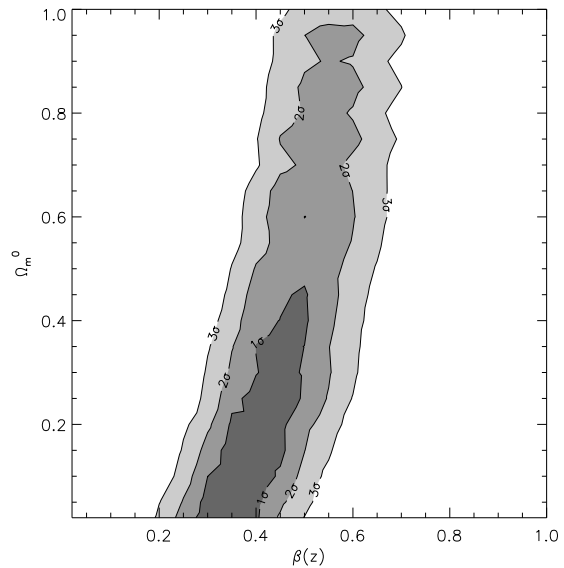,width=8cm}}
\caption[Likelihood contours of $\Omega_{\rm m}^{0}$-$\beta(z=0.55)$]
{Likelihood contours of $\Omega_{\rm m}^{0}$-$\beta(z=0.55)$ using the 
geometric method of Alcock-Paczynski test and modelling the redshift-space
distortions. The best-fit values are $\Omm=0.10^{+0.35}_{-0.10}$ and
$\beta(z=0.55)=0.40\pm0.05$ with a velocity dispersion of
$\sigma=330$kms$^{-1}$. 
Note how a value of $\Omega_{\rm m} \sim 0.3$ is not ruled out but
also the large degeneracy along the $\Omega_{\rm m}$ direction. 
A $\Lambda$ cosmology is assumed, along with a model where 
$\gamma = 1.72$ and a (starting) value of $r_{0} = 7.45 \hmpc$.}
\label{fig:ppt1}
\end{figure}

Details on the fitting procedure are given in dA05 (Section 7.7). Using
this AP-distortion test, we calculate values of $\Omega_{\rm m}$-$\beta$
for the assumed $\Lambda$ cosmology.
and present them in  Figure~\ref{fig:ppt1}. We first note that the
constraint here is almost entirely on $\beta$ rather than $\Omm$.
Using the $\xi(r)$ fit with a (starting) $r_0=7.45 \hmpc$ and
$\gamma= 1.72$, we find that $\Omm=0.10^{+0.35}_{-0.10}$ and
$\beta(z=0.55)=0.40\pm0.05$ with a velocity dispersion of
$\sigma=330$kms$^{-1}$ from a $\chi^{2}$ minimization. 
We have checked these errors by repeating the above calculations on the 32
``jackknife'' sub-samples. In order to make the jackknife calculations less
computationally intensive, the velocity dispersion is held fixed at
330$\kms$ in every case. 
Comparing the error contours in Fig.~\ref{fig:ppt1} with the jackknife 
estimates, we again find that the jackknife errors for $\beta$ at 
$\pm0.05$ are comparable to, if not smaller than, those in the error 
contours in Fig.~\ref{fig:ppt1}. The jackknife error in $\Omm$ at $\pm0.14$ 
is comparable to the error contour in Fig.~\ref{fig:ppt1}. 
As in Fig.~\ref{fig:fit_s0_gamma_full_rv1_xir_LS_jacks_paper}, this
agreement may be surprising given that we have ignored the covariance in
$\xi(\sigma,\pi)$ points which is almost certainly non-negligible. Again we 
argue that a relatively poor $\chi^{2}$ fit at minimum may be responsible, 
leading to a somewhat fortuitous agreement of the formal and jackknife error.
But on the grounds of the jackknife results we believe that the error 
contours shown in Fig.~\ref{fig:ppt1} are reasonably realistic and 
we shall quote these hearafter. 

\begin{figure}
\centering
\centerline{\psfig{file=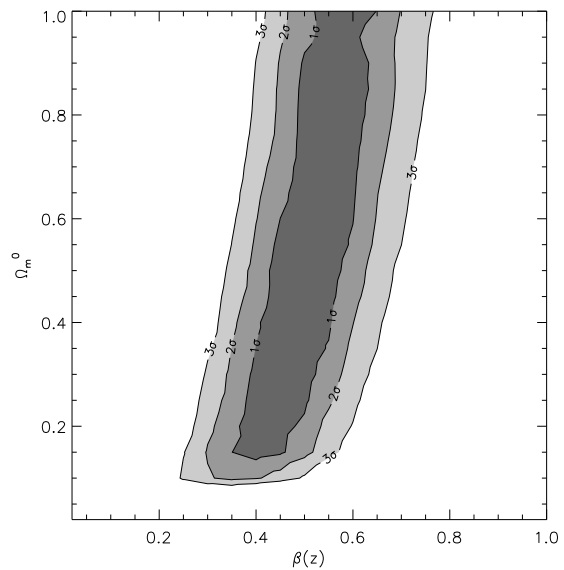,width=8cm}}
\caption{Likelihood contours of $\Omega_{\rm m}^{0}$-$\beta(z=0.55)$ using 
the geometric method of Alcock-Paczynski test and modelling the redshift-space
distortions, assuming an EdS cosmology. The best-fit values are 
$\Omm=0.40^{+0.60}_{-0.25}$ and $\beta=0.45^{+0.20}_{-0.10}$ using a model 
with $\gamma = 1.67$ and a (starting) correlation length of $r_{0}=5.65\hmpc$. 
A value of $\Omega_{\rm m} \sim 1.0$ lies within our 1$\sigma$ contour 
but again there is a large degeneracy along the $\Omega_{\rm m}$ direction.}
\label{fig:ppt1_EdS}
\end{figure}

We have also fitted $\xisp$ assuming an EdS cosmology.  In principle
this  should give the same result as assuming the $\Lambda$ model.  We
show these $\Omm-\beta$ fits in Figure~\ref{fig:ppt1_EdS}.  We find
that the best fit is now $\Omm=0.40^{+0.6}_{-0.25}$ and
$\beta=0.45^{+0.20}_{-0.10}$ ($\chi^{2}$ minimization) with a  velocity
dispersion of $\sigma=330$kms$^{-1}$.  A model with $\gamma = 1.67$ and
a (starting) correlation length of  $r_{0}=5.65\hmpc$ is used.  Thus
the $\beta$ and the velocity dispersion values are reasonably
consistent with the previous result. However, the value of $\Omm$
assuming  an EdS cosmology, is somewhat higher than the best-fit found
assuming a $\Lambda$ cosmology.  We assume that the high degeneracy of
$\Omm$ coupled with slightly different $\xi(r)$ models in the two cases
is causing this discrepancy. The contours in Fig.~\ref{fig:ppt1_EdS}
certainly suggest that the constraint  on $\Omm$ is much less strict in
the EdS assumed case.

We have investigated other systematics in the $\Omm-\beta$ fits.
Returning to an assumed $\Lambda$ cosmology, there is some small 
dependence on the model assumed for $\xi(r)$. 
For example, if the slope $\gamma=1.69$ from fitting
$\xi(r)$ in the more limited range $0.4<r<20$h$^{-1}$Mpc is assumed then
we find that $\Omm=0.10\pm0.29$ and $\beta(z=0.55)=0.35\pm0.16$ with a
velocity dispersion of $\sigma=300$kms$^{-1}$. 
Further, if instead of using $\xi(r)$, $w_{p}(\sigma)$ is used with slope 
$\gamma=1.83$ over the usual $0.4 < r < 50 \hmpc$ range, 
we find that the best-fit model prefers a very low value of
$\Omm=0.02\pm0.15$ and $\beta(z=0.55)=0.40\pm0.05$ 
with a velocity dispersion of $\sigma=360$kms$^{-1}$. 
The consistency of these different models to give values of
$\Omm, \beta$ and a pairwise velocity dispersion, albeit at a cost of
a very loose constraint on $\Omm$, 
is re-assuring and summarised in Table~\ref{tab:zspace_dist_values}.
Since  $w(\theta)$ also seems to indicate a flatter 
($\gamma=-1.67\pm{0.07}$) slope in the $1<r<20 \hmpc$ range of interest
for $\xi(\sigma,\pi)$ we take our `best bet' estimates to be the values
for $\gamma=-1.72$ given above. 
These values also give a good overall fit 
to $\xi(s)$. We next introduce a further constraint to break the 
$\Omm-\beta$ degeneracy.
\begin{table*}
\centering
\caption{Best fitting model values of $\Omm, \beta$ and 
pairwise velocity dispersion, $\wrms$, 
using redshift-space distortions alone and assuming a $\Lambda$ cosmology.}
\begin{tabular}{||l|c|c|c|l|c|c} \hline
\hline
$r_{0}$ & $\gamma$ &  range /$\hmpc$ & Measure & $\Omm$ & $\beta$ & $\wrms / \kms$ \\
\hline
7.45     &  1.72    &  0.4-50        & $\xir$  &   0.10 &  0.40   &  330    \\
7.30     &  1.83    &  0.4-50	     & $\wp $  &   0.02 &  0.40   &  360    \\
7.60     &  1.68    &  0.4-20        & $\xir$  &   0.10 &  0.35   &  300    \\
7.34     &  1.80    &  0.4-20        & $\wp$   &   0.10 &  0.45   &  360    \\
\hline
\end{tabular}
\label{tab:zspace_dist_values}
\end{table*}

\subsection{Further Constraints on $\Omega_{0}^{\rm m}$ and $\beta(z)$
from LRG Clustering Evolution}

\citet{Matsubara96} and \citet{Croom96} pointed out that by
combining low redshift and high redshift clustering information, further
constraints on $\Omm$ and $\Omlam$ would be possible. The basic idea
described in this section is that the $\Omega_{\rm m}$:$\beta(z)$ 
degenerate set obtained from LRG clustering evolution is different from
the $\Omega_{\rm m}$:$\beta(z)$ degenerate set obtained from analysing
LRG redshift-space distortions; by using these two constraints in
combination, the degeneracies may be lifted. Thus the way we proceed to
break the degeneracy is to combine our current 2SLAQ LRG results with
constraints derived from consideration of LRG clustering evolution. 

From the value of the mass correlation function at $z=0$, linear
perturbation theory can be used, assuming a test $\Omm$, to compute
the value of the mass correlation function in real space at $z=0.55$.
This can then be compared to the measured LRG $\xi(r)$ at $z=0.55$ to
find the value of the bias $b(z=0.55)$. The clustering of the mass at
$z=0$ can be determined if the galaxy correlation function is known,
assuming that the bias of the galaxies used, $b(z=0)$, is independent
of scale. Fortunately, recent galaxy redshift surveys have obtained
precise measurements of the clustering of galaxies at $z \approx
0$. In practice we shall start from $\xis$ at $z=0$ and $z=0.55$ and
use equation~\ref{eq:xis_div_xir} to derive $\xi(r)$ in each case. 

We therefore follow \citet[]{daAngela05} and start by introducing the
volume averaged two-point correlation function $\bar{\xi}$ where
\begin{equation}
\bar{\xi}(s) = \frac{\int_{0}^{s} 4\pi s'^{2} \xi(s') ds}
                 {\int_{0}^{s} 4\pi s'^{2} ds}
\label{eq:xi_bar_jaca}
\end{equation}
We do this so that non-linear effects in the sample should be
insignificant due to the $s^{2}$ weighting, setting the upper limit of
the integral $s=20\hmpc$.  To calculate equation~\ref{eq:xi_bar_jaca}
at $z=0$, we use the  double-power law form that is found by the 2dFGRS
to  describe $\xis$ \citep[][Fig. 6]{Hawkins03} in the numerator.

Then, the equivalent averaged correlation function in real-space
can be determined by 
\begin{equation}
\bar{\xi}(r,z=0) = \frac{\bar{\xi}(s,z=0)}
                 {1 + \frac{2}{3} \beta (z=0) + \frac{1}{5} \beta(z=0)^{2}}
\end{equation}
where $\bar{\xi}(s)$ comes from equation~\ref{eq:xi_bar_jaca} and
we take the value of $\beta$ for the 2dFGRS as $\beta(z=0) = 0.49\pm0.09$
\citep{Hawkins03}. 
Now the real-space mass correlation is obtained with
\begin{equation}
\bar{\xi}_{\rm mass} (r,z=0) = \frac{\bar{\xi}(r,z=0)}
                                   {b(z=0)^{2}},
\end{equation}
where $b(z=0)$ is given for each test cosmology by 
\begin{equation}
b(z=0) = \frac{\Omega^{0.6}_{\rm m}(z=0)}
              {\beta(z=0)}.
\end{equation}

Once we have
determined the real-space correlation function of the mass at  $z=0$,
its value at $z=0.55$ is obtained using linear perturbation theory.
Hence, at  $z=0.55$, the real-space correlation function of the mass
will be:
\begin{equation}
\bar{\xi}_{\rm mass} (r,z=0.55) = \frac{\bar{\xi}_{\rm mass} (r,z=0)}
                                         {G(z=0.55)^{2}},
\label{equation:bias_mass0.55}
\end{equation}
Here, $\bar{\xi}_{\rm mass}(r)$ is the volume-averaged correlation
function (with $1 \leq r \leq 20 \hmpc$) and $G(z)$ is the growth
factor of perturbations, given by linear theory \citep{Carroll92,
Peebles80} and depends on cosmology - in this case the test cosmology.

Once the value of $\bar{\xi}_{\rm mass} (r,z=0.55)$ is obtained for a given 
test cosmology, the process to find $\beta(z=0.55)$ is similar to the one used to 
find $\bar{\xi}_{\rm mass} (r,z=0)$, but now the steps are performed in reverse:
$\bar{\xi}(s,z=0.55)$ can be measured in a similar way as 
$\bar{\xi}(s,z=0)$. The bias factor at $z\approx 0.55$ is given by:
\begin{equation}
b^{2}(z=0.55) = \frac{\bar{\xi}(r,z=0.55)}
                     {\bar{\xi}_{\rm mass} (r,z=0.55)},
\label{equation:bias_z0.55}
\end{equation}
where $\bar{\xi}_{\rm mass}(r)$ is given by equation \ref{equation:bias_mass0.55} 
and $\bar{\xi}(r,z=0.55)$ is obtained by:
\begin{equation}
\bar{\xi}(r,z=0.55) = \frac{\bar{\xi}(s,z=0.55)}
                      {1+\frac{2}{3}\beta(z=0.55)+\frac{1}{5}\beta(z=0.55)^{2}}.
\label{equation:kaiser_average2}
\end{equation}
The value of $\beta(z=0.55)=\Omega_{\rm m}^{0.6}(z=0.55)/b(z=0.55)$ 
will be obtained, for the given test value of $\Omega_{\rm m}(z=0)$ 
by solving the second order polynomial formed by substituting 
$\bar{\xi}(r,z=0.55)$ from equation 40 into equation 41 
\citep[see][]{Hoyle02}. The confidence levels on the computed values of
$\beta(z=0.55)$ are calculated by combining appropriately in quadrature
the errors on $\bar{\xi}(s,z=0.55)$, $\bar{\xi}(s,z=0)$, $\beta(z=0.55)$ 
and $\beta(z=0)$.

\begin{figure}
\centering
\centerline{\psfig{file=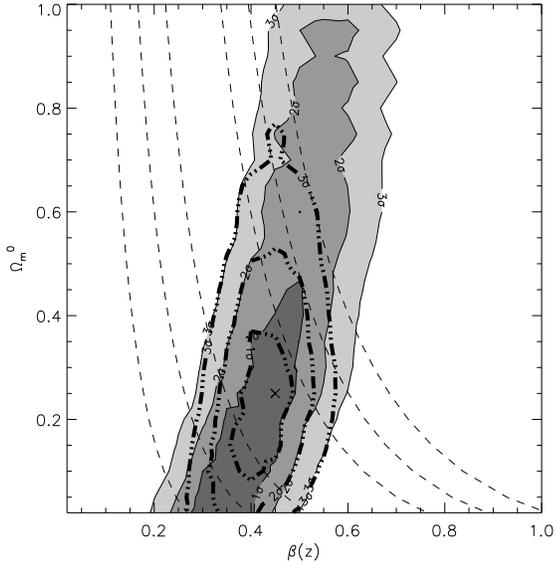,width=8cm}}
\caption{Joint likelihood contours of $\Omega_{\rm m}^{0}$-$\beta(z=0.55)$ 
using the geometric method of Alcock-Paczynski test, modelling the 
redshift-space distortions and including the evolution of clustering 
constraints, assuming the $\Lambda$ cosmology. Here we see that the best-fit
joint constraint values are $\Omega_{\rm m} = 0.25^{+0.10}_{-0.15}$, 
$\beta = 0.45\pm0.05$ (marked with the cross) with a $\wrms$ of 
$330 \rm kms^{-1}$.}
\label{fig:ppt3}
\end{figure}

\begin{figure}
\centering
\centerline{\psfig{file=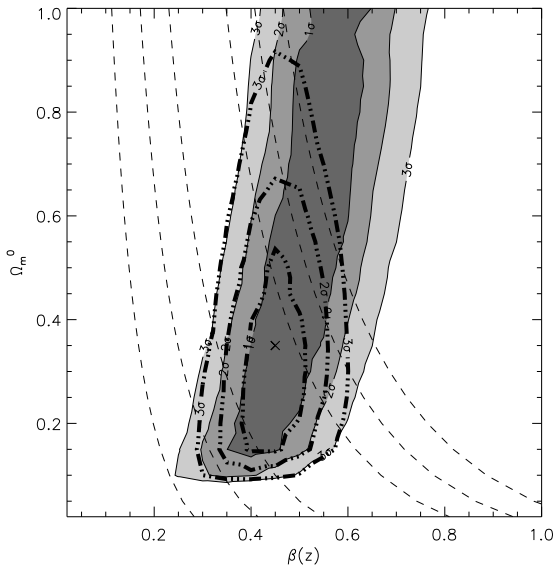,width=8cm}}
\caption{Joint likelihood contours for $\Omega_{\rm m}^{0}$-$\beta(z=0.55)$ 
using the geometric method of Alcock-Paczynski test, modelling the 
redshift-space distortions and including the evolution of clustering 
constraints, assuming an EdS cosmology. 
The joint best-fit has $\Omm=0.35\pm0.15$, $\beta=0.45\pm0.05$ (marked with
the cross) and a $\wrms$ of $330 \rm kms^{-1}$. When the joint constraints are
considered, a value of $\Omm =1.0$ can be ruled out at the 3$\sigma$ level.}
\label{fig:ppt3_EdS}
\end{figure}

Combining this clustering evolution constraint with those 
from $z$-space distortions breaks the degeneracy in the 
$\Omega_{\rm m}-\beta$ plane. We can now work out
the joint-2 parameter best fitting regions. This is shown in
Figure~\ref{fig:ppt3}, where the 1, 2 and 3 sigma error bars  are
plotted (dashed lines). The best fitting 2-parameter calculations has
$\Omega_{\rm m} = 0.25$, $\beta = 0.45$ denoted by the cross in
Figure~\ref{fig:ppt3}. When we consider the 1-sigma error on each
quantity separately we find, $\Omega_{\rm m} = 0.25^{+0.10}_{-0.15}$,
$\beta = 0.45\pm0.05$ with a $\wrms$ of $330 \rm kms^{-1}$.  A model
$\xi(r)$ is assumed with $\gamma=1.72$ and $r_{0}=7.45 \hmpc$,  as is a
$\Lambda$ cosmology.

The case of the combined constraint for the EdS assumed cosmology is
shown in Figure~\ref{fig:ppt3_EdS}. The $\xi(r)$ model with
$\gamma=1.67$ and $r_{0}=5.65 \hmpc$ is assumed and we find
$\Omm=0.35\pm0.15$ and $\beta=0.45\pm0.05$. Although the 3-sigma
contours still reject the EdS model, the rejection is less than in the
$\Lambda$ assumed case. Overall we conclude that the combined
constraints on $\beta$ are the strongest with $\beta=0.45\pm0.05$
consistently produced whatever the assumed cosmology or $\xi(r)$
model. Though the combined constraints on $\Omm$ are less strong and
give $\Omm\approx0.3\pm0.15$, they still appear consistent with the
standard $\Lambda$ model.

\begin{figure}
\centering \centerline{\psfig{file=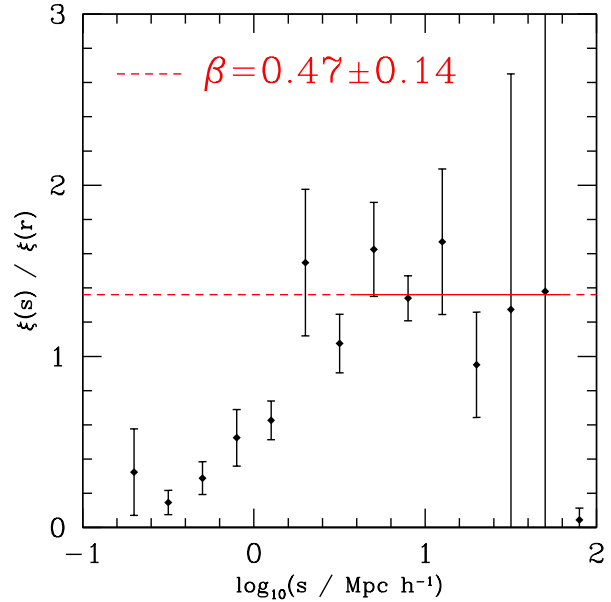,width=8cm}}
\caption[The ratio of the redshift-space correlation function to
the real-space correlation function, measured from the 2SLAQ LRG
survey.]
{The ratio of the redshift-space correlation function to
the real-space correlation function, measured from the 2SLAQ LRG
survey. We assume a $\Lambda$ cosmology for these measurements and
fitting over the scales of 5 - 50 $\hmpc$ find that $\beta= 0.47\pm0.14$, 
in very good agreement with our redshift-space distortion/evolution of
clustering technique measurements.}
\label{fig:xis_div_xir_beta}
\end{figure}

As another check, we can use the ratio $\xis / \xir$ to determine
$\beta$ from equation~\ref{eq:xis_div_xir} (see
Figure~\ref{fig:xis_div_xir_beta}). We assume that $\beta$ is
scale-independent, the $z$-space distortions are only affected by the
large-scale infall and are not contaminated by random peculiar
motions. Fitting over the scales, $ 5 < s < 50 \hmpc$,  we find
$\beta=0.47\pm0.14$, which is consistent with our determination using
the distortions. The 1-$\sigma$ error comes from a standard $\chi^{2}$
analysis using the $\xi(s)/\xi(r)$ ratios and their errors; these are
derived from adding the jackknife errors on $\xi(s)$ and $\xi(r)$ in
quadrature. We note that this procedure does not take into account the
non-independence of the correlation  function points, suggesting that
the relatively large error quoted above on $\beta$ may still be a
lower limit.

The low values of $\Omm \approx 0.30$ and the value of $\beta=0.45$ we
find from the 2SLAQ LRG survey are in line with what is generally
expected in the current standard cosmological model. Although the
constraint on $\beta$ is tight, the constraint on $\Omm$ is less so and
in particular the EdS value is not rejected  at 3$\sigma$ when
clustering distortions only are considered. However, when the combined
evolution and redshift distortions are considered, the EdS value is
rejected at the 3$\sigma$ level.

Using equations~\ref{equation:bias_z0.55}, \ref{equation:kaiser_average2},
$\Omm(z=0)=0.30\pm0.15$ and $\beta(z=0.55)=0.45\pm0.05$, we find that
$b(z=0.55) = \Omega^{0.6}_{\rm m}(z=0.55)/\beta(z=0.55) =
1.66\pm0.35$, showing that the 2SLAQ LRGs are highly biased objects.
This can be compared with the value for SDSS LRGs at redshift $z=0.55$
which are found to have a value of $b=1.81 \pm 0.04$ (Padmanabhan et
al. 2006, Fig. 13). The 2SLAQ LRG value is consistent with this SDSS
LRG value; of course a slightly lower bias may have been expected for
2SLAQ LRGs due to the bluer/lower luminosity selection cut.  If we
assume the value found in recent studies of $\Omm(z=0)=0.25$
\citep{Cole05, Eisenstein05, Tegmark06, Percival06a, Percival06b},
then our esimate of $b$ becomes $b=1.56\pm0.33$.

Although we leave discussion about the bias estimate and the accuracy
of the  $\beta$-model to a future paper, at the referee's request, we
compare  the non-linear mass correlation function as numerically
calculated for the  standard cosmology \citep{Colin99} to the 2SLAQ LRG
$\xi(r)$,  in Figure~\ref{fig:xir_2SLAQ_Cov_EdS_div_plfit}.   The
errors in $\xi(\sigma,\pi)$ are smaller at separations 5 to 20 $\hmpc$,
than at $1 \hmpc$, so our estimates of bias from $\xi(\sigma,\pi)$ are
weighted towards these larger scales where there appears to be
approximate consistency with the relative amplitudes of $\xi_{\rm
mass}$ and $\xi(r)$ in Fig.~\ref{fig:xir_2SLAQ_Cov_EdS_div_plfit}.
Thus, as mentioned previously, our working assumption from here on will
be that there is no effect of scale-dependent bias  on our $\Omega_{\rm
m}-\beta$ fits and we leave further investigation of this issue for 
future work.

Finally, taking the value of $b(z=0.55)=1.66\pm0.35$, we can relate
$b(z=0)$ to $b(0.55)$ using the bias evolution model \citep{Fry96}
\begin{equation}
b(z) = 1 + [b(0) - 1] G(\Omm(0), \Omlam(0), z), 
\label{eqn:bias_model_1}
\end{equation}
where $G(\Omm(0), \Omlam(0), z)$ is the linear growth rate of the
density perturbations \citep{Peebles80, Peebles84, Carroll92}.  There
are many other bias models, but here we are making the simple
assumptions that galaxies formed at early times and their subsequent
clustering is governed purely by their discrete motion within the
gravitational potential produced by the matter density perturbations.
This model would be appropriate, for example, in a  ``high-peaks''
biasing scenario where early-type galaxies formed at a single  redshift
and their co-moving space density then remained constant to the present
day. There may be evidence for such a simple evolutionary history in
the observed early-type stellar mass/luminosity functions
\citep[e.g.][]{Metcalfe01, Brown06, Wake06}.  From
equation~\ref{eqn:bias_model_1}, and taking $b(0.55)=1.66$,  implies a
value today of $b(0)=1.52$ at $z\sim0.1$.  This leads to a predicted
correlation length today  of $r_{0}(z=0) = 8.5\pm1.6 \hmpc$ (assuming
$\Lambda$CDM) 
which is consistent with the 2dFGRS value of $r_{0}=8.0\pm1.0 \hmpc$ found
from averaging the same two matched luminosity bins from Table 2 of
\citet{Norberg02a}, and previously used in our Fig. 7. (But note that 
the 2dFGRS $\xis$ shown in Fig. 7 might imply a somewhat lower value for the
2dFGRS clustering amplitude in this bin than $r_{0}=8.0\pm1.0 \hmpc$.)

Therefore, these correlation function evolution results suggest that
there  seems to be no inconsistency with the idea that the LRGs have a
constant  co-moving space density, as may be suggested by the
luminosity function  results.  But, we note that the LF results of
\citet{Wake06} apply to a colour-cut  sample, (where 2SLAQ LRGs are
carefully matched to SDSS LRGs) whereas our  clustering results are
only approximately matched to the 2dFGRS.  It will be interesting to
see if this results holds when the clustering  of the exactly matched
high and low redshift LRGs are compared  (see Wake et al., 2007, in
prep.).

%%%%%%%%%%%%%%%%%%%%%%%%%%%%%%%%%%%%%%%%%%%%%%%%%%%%%%%%%%%%%%%%%%%%
%SECTION 5  SECTION 5  SECTION 5  SECTION 5  SECTION 5  SECTION 5  %
%SECTION 5  SECTION 5  SECTION 5  SECTION 5  SECTION 5  SECTION 5  %
%SECTION 5  SECTION 5  SECTION 5  SECTION 5  SECTION 5  SECTION 5  %
%%%%%%%%%%%%%%%%%%%%%%%%%%%%%%%%%%%%%%%%%%%%%%%%%%%%%%%%%%%%%%%%%%%%
\section{conclusions}\label{sec:conclusions}

We have performed a detailed analysis of the clustering of 2SLAQ LRGs in
redshift space as described by the two-point correlation function.
Our main conclusions are as follows.
 
\begin{enumerate}
\item{The LRG two-point correlation function, $\xi(s)$,
averaged over the redshift range $0.4<z<0.8$, shows a slope which
changes as a function of scale, being flatter on small scales and
steeper on large scales, consistent with the expected effects of 
redshift-space distortions.}

\item{The best fitting single power-law model to the 
Real-space 2-Point correlation function of the 2SLAQ LRG Survey has
a clustering length of $r_{0} = 7.45\pm0.35 \hmpc$ and a power-law slope of 
$\gamma = 1.72\pm0.06$ (assuming a $\Lambda$ cosmology) 
showing LRGs to be highly clustered objects.}

\item{Evidence for a change in the slope of the projected 
correlation function, which is a prediction of halo occupation 
distribution (HOD) models, is seen in the 2SLAQ LRG survey results, while
a stronger feature is observed in the angular correlation function of the 
LRGs. A direct explanation for this remains unclear.} 

\item{From redshift distortion models and the geometric Alcock-Paczynski
test we find $\Omm=0.10^{+0.35}_{-0.10}$ and $\beta(z=0.55)=0.40\pm0.05$ 
with a velocity dispersion of $\sigma=330$kms$^{-1}$, assuming a $\Lambda$ 
cosmology.  With EdS as the assumed cosmology, $\Omm=0.40^{+0.60}_{-0.25}$ 
and $\beta=0.45^{+0.20}_{-0.10}$ with the best-fitting velocity dispersion 
remaining at $\sigma=330$kms$^{-1}$. 
However, in both cases, we also find a degeneracy along the 
$\Omega_{\rm mass,0}$-$\beta$ plane.} 

\item{By considering the evolution of clustering from $z \sim 0$ to
$z_{\rm LRG} =0.55$ we can break this degeneracy and find that
$\Omega_{\rm m} = 0.25^{+0.10}_{-0.15}$ and $\beta = 0.45\pm0.05$ 
(with a $\wrms$ of $330 \rm kms^{-1}$) assuming a $\Lambda$ cosmology.
When the EdS cosmology is assumed, we find $\Omm=0.35\pm0.15$ and 
$\beta=0.45\pm0.05$ (again $\wrms = 330 \rm kms^{-1}$). When the joint
constraints are considered, a value of $\Omm=1.0$ can be ruled out
at the 3$\sigma$ level.
We believe these estimates of $\beta(z=0.55)$ are reasonably robust but
the values of $\Omm$ are less well constrained, although the above estimate
for $\Omm=0.30\pm{0.15}$ is in agreement with concordance values.}

\item{If we assume a $\Lambda$ cosmology with $\Omm(z=0)=0.3$ and 
$\beta(z=0.55)=0.45$ then the value for the 2SLAQ LRG bias at $\bar{z}=0.55$ 
is $b=1.66\pm0.35$, in line with other recent measurements of LRG bias 
(Padmanabhan et al. 2006).} % ``Point (vi)''

\item{Assuming this $b(z=0.55)=1.66$ value, and adopting a simple 
``high-peaks'' bias prescription which assumes LRGs have a constant co-moving
space density, we predict $r_{0}=8.5\pm1.6 \hmpc$ for LRGs at $z \approx 0.1$.
This is not inconsistent with the observed result for luminosity matched 
2dFGRS `LRGs' at this redshift.}

\end{enumerate}

The clustering and redshift-space distortion results complement 
the other results from the 2SLAQ Survey e.g. \citet{Wake06}, 
Wake et al. (2007, in prep.) and da Angela et al. (2006, in prep).
Luminous Red Galaxies may be considered to be ``red and dead'' but they
have recently been realised to be very powerful tools for both constraining 
galaxy formation and evolution theories as well as cosmological probes. 
Future projects utilising LRGs 
(e.g. to measure the baryon acoustic oscillations 
or to study LRGs at higher redshift/fainter magnitudes)
will give us more insights into today's greatest astrophysical problems, 
including the epoch of massive galaxy formation and the acceleration of 
the cosmological expansion.

%%%%%%%%%%%%%%%%%%%%%%%%%%%%%%%%%%%%%%%%%%%%%%%%%%%%%%%%%%%%%%%%%%%%
%SECTION 6  SECTION 6  SECTION 6  SECTION 6  SECTION 6  SECTION 6  %
%SECTION 6  SECTION 6  SECTION 6  SECTION 6  SECTION 6  SECTION 6  %
%SECTION 6  SECTION 6  SECTION 6  SECTION 6  SECTION 6  SECTION 6  %
%%%%%%%%%%%%%%%%%%%%%%%%%%%%%%%%%%%%%%%%%%%%%%%%%%%%%%%%%%%%%%%%%%%%
%%%%%%%%%%%%%%%%%%%%%%%%%%%%%%%%%%%%%%%%%%%%%%%%%%%%%%%%%%%%%%%%%%%%
%SECTION 7  SECTION 7  SECTION 7  SECTION 7  SECTION 7  SECTION 7  %
%SECTION 7  SECTION 7  SECTION 7  SECTION 7  SECTION 7  SECTION 7  %
%SECTION 7  SECTION 7  SECTION 7  SECTION 7  SECTION 7  SECTION 7  %
%%%%%%%%%%%%%%%%%%%%%%%%%%%%%%%%%%%%%%%%%%%%%%%%%%%%%%%%%%%%%%%%%%%%
%%%%%%%%%%%%%%%%%%%%%%%%%%%%%%%%%%%%%%%%%%%%%%%%%%%%%%%%%%%%%%%%%%%%
%SECTION 8  SECTION 8  SECTION 8  SECTION 8  SECTION 8  SECTION 8  %
%SECTION 8  SECTION 8  SECTION 8  SECTION 8  SECTION 8  SECTION 8  %
%SECTION 8  SECTION 8  SECTION 8  SECTION 8  SECTION 8  SECTION 8  %
%%%%%%%%%%%%%%%%%%%%%%%%%%%%%%%%%%%%%%%%%%%%%%%%%%%%%%%%%%%%%%%%%%%%
%%%%%%%%%%%%%%%%%%%%%%%%%%%%%%%%%%%%%%%%%%%%%%%%%%%%%%%%%%%%%%%%%%%%
%%%%%%%%%%%%%%%%%%%%%%%%%%%%%%%%%%%%%%%%%%%%%%%%%%%%%%%%%%%%%%%%%%%%
\section*{acknowledgements}
NPR acknowledges a PPARC Studentship 
and J. da \^{A}ngela acknowledges financial support from FCT/Portugal through
project POCTI/FNU/43753/2001 and also ESO/FNU/43753/2001.
We thank P. Norbreg, J. Tinker, S. Cole and C. Baugh as well as the 
referee for stimulating discussion and useful comments.
We warmly thank all the present and former staff of the
Anglo-Australian Observatory for their work in building and operating
the 2dF facility.  The 2SLAQ Survey is based on
observations made with the Anglo-Australian Telescope and for the SDSS. 
Funding for the creation and distribution of the SDSS Archive has been 
provided by the Alfred P. Sloan Foundation, the Participating Institutions, 
the National Aeronautics and Space Administration, the National Science 
Foundation, the U.S. Department of Energy, the Japanese Monbukagakusho, 
and the Max Planck Society. The SDSS Web site is http://www.sdss.org/.
The SDSS is managed by the Astrophysical Research Consortium (ARC) for the 
Participating Institutions. The Participating Institutions are The University 
of Chicago, Fermilab, the Institute for Advanced Study, the Japan Participation
Group, The Johns Hopkins University, the Korean Scientist Group, Los Alamos 
National Laboratory, the Max-Planck-Institute for Astronomy (MPIA), 
the Max-Planck-Institute for Astrophysics (MPA), New Mexico State University, 
University of Pittsburgh, University of Portsmouth, Princeton University, 
the United States Naval Observatory, and the University of Washington.

\bibliographystyle{mn2e}
\bibliography{tester_mnras}

\begin{thebibliography}{}

\bibitem[\protect\citeauthoryear{{Alcock} \& {Paczynski}}{{Alcock} \&
  {Paczynski}}{1979}]{AP79}
{Alcock} C.,  {Paczynski} B.,  1979, \nat, 281, 358

\bibitem[\protect\citeauthoryear{{Ballinger}, {Peacock} \&
  {Heavens}}{{Ballinger} et~al.}{1996}]{Ballinger96}
{Ballinger} W.~E.,  {Peacock} J.~A.,    {Heavens} A.~F.,  1996, \mnras, 282,
  877

\bibitem[\protect\citeauthoryear{{Blanton}, {Eisenstein}, {Hogg} \&
  {Zehavi}}{{Blanton} et~al.}{2006}]{Blanton06}
{Blanton} M.~R.,  {Eisenstein} D.,  {Hogg} D.~W.,    {Zehavi} I.,  2006, \apj,
  645, 977

\bibitem[\protect\citeauthoryear{{Brown}, {Dey}, {Jannuzi}, {Brand}, {Benson},
  {Brodwin}, {Croton} \& {Eisenhardt}}{{Brown} et~al.}{2006}]{Brown06}
{Brown} M.~J.~I.,  {Dey} A.,  {Jannuzi} B.~T.,  {Brand} K.,  {Benson} A.~J.,
  {Brodwin} M.,  {Croton} D.~J.,    {Eisenhardt} P.~R.,  2006, pre-print
  (astro-ph/0609584)

\bibitem[\protect\citeauthoryear{{Cannon} et~al.,}{{Cannon}
  et~al.}{2006}]{Cannon06}
{Cannon} R.,  et~al., 2006, \mnras, 372, 425

\bibitem[\protect\citeauthoryear{{Carroll}, {Press} \& {Turner}}{{Carroll}
  et~al.}{1992}]{Carroll92}
{Carroll} S.~M.,  {Press} W.~H.,    {Turner} E.~L.,  1992, \araa, 30, 499

\bibitem[\protect\citeauthoryear{{Coil} et~al.,}{{Coil}  et~al.}{2004}]{Coil04}
{Coil} A.~L.,  et~al., 2004, \apj, 609, 525

\bibitem[\protect\citeauthoryear{{Cole} et~al.,}{{Cole}  et~al.}{2005}]{Cole05}
{Cole} S.,  et~al., 2005, \mnras, 362, 505

\bibitem[\protect\citeauthoryear{{Coles} \& {Erdogdu}}{{Coles} \&
  {Erdogdu}}{2007}]{Coles07}
{Coles} P.,  {Erdogdu} P.,  2007, astro-ph/0706.0412, 706

\bibitem[\protect\citeauthoryear{{Col{\'{\i}}n}, {Klypin}, {Kravtsov} \&
  {Khokhlov}}{{Col{\'{\i}}n} et~al.}{1999}]{Colin99}
{Col{\'{\i}}n} P.,  {Klypin} A.~A.,  {Kravtsov} A.~V.,    {Khokhlov} A.~M.,
  1999, \apj, 523, 32

\bibitem[\protect\citeauthoryear{{Croom} et~al.,}{{Croom}
  et~al.}{2005}]{Croom05}
{Croom} S.~M.,  et~al., 2005, \mnras, 356, 415

\bibitem[\protect\citeauthoryear{{Croom} \& {Shanks}}{{Croom} \&
  {Shanks}}{1996}]{Croom96}
{Croom} S.~M.,  {Shanks} T.,  1996, \mnras, 281, 893

\bibitem[\protect\citeauthoryear{{Croom}, {Smith}, {Boyle}, {Shanks}, {Miller},
  {Outram} \& {Loaring}}{{Croom} et~al.}{2004}]{Croom04}
{Croom} S.~M.,  {Smith} R.~J.,  {Boyle} B.~J.,  {Shanks} T.,  {Miller} L.,
  {Outram} P.~J.,    {Loaring} N.~S.,  2004, \mnras, 349, 1397

\bibitem[\protect\citeauthoryear{{da Angela} et~al.,}{{da Angela}
  et~al.}{2006}]{daAngela06}
{da Angela} J.,  et~al., 2006, astro-ph/0612401

\bibitem[\protect\citeauthoryear{{da {\^A}ngela}, {Outram}, {Shanks}, {Boyle},
  {Croom}, {Loaring}, {Miller} \& {Smith}}{{da {\^A}ngela}
  et~al.}{2005}]{daAngela05}
{da {\^A}ngela} J.,  {Outram} P.~J.,  {Shanks} T.,  {Boyle} B.~J.,  {Croom}
  S.~M.,  {Loaring} N.~S.,  {Miller} L.,    {Smith} R.~J.,  2005, \mnras, 360,
  1040

\bibitem[\protect\citeauthoryear{{Davis} \& {Peebles}}{{Davis} \&
  {Peebles}}{1983}]{Davis83}
{Davis} M.,  {Peebles} P.~J.~E.,  1983, \apj, 267, 465

\bibitem[\protect\citeauthoryear{{Eisenstein} et~al.,}{{Eisenstein}
  et~al.}{2001}]{Eisenstein01}
{Eisenstein} D.~J.,  et~al., 2001, \aj, 122, 2267

\bibitem[\protect\citeauthoryear{{Eisenstein} et~al.,}{{Eisenstein}
  et~al.}{2005}]{Eisenstein05}
{Eisenstein} D.~J.,  et~al., 2005, \apj, 633, 560

\bibitem[\protect\citeauthoryear{{Fry}}{{Fry}}{1996}]{Fry96}
{Fry} J.~N.,  1996, \apjl, 461, L65

\bibitem[\protect\citeauthoryear{{Fukugita}, {Ichikawa}, {Gunn}, {Doi},
  {Shimasaku} \& {Schneider}}{{Fukugita} et~al.}{1996}]{Fukugita96}
{Fukugita} M.,  {Ichikawa} T.,  {Gunn} J.~E.,  {Doi} M.,  {Shimasaku} K.,
  {Schneider} D.~P.,  1996, \aj, 111, 1748

\bibitem[\protect\citeauthoryear{{Hamilton}}{{Hamilton}}{1993}]{Hamilton93}
{Hamilton} A.~J.~S.,  1993, \apj, 417, 19

\bibitem[\protect\citeauthoryear{{Hatton} \& {Cole}}{{Hatton} \&
  {Cole}}{1998}]{Hatton98}
{Hatton} S.,  {Cole} S.,  1998, \mnras, 296, 10

\bibitem[\protect\citeauthoryear{{Hawkins} et~al.,}{{Hawkins}
  et~al.}{2003}]{Hawkins03}
{Hawkins} E.,  et~al., 2003, \mnras, 346, 78

\bibitem[\protect\citeauthoryear{{Hoyle}, {Outram}, {Shanks}, {Boyle}, {Croom}
  \& {Smith}}{{Hoyle} et~al.}{2002}]{Hoyle02}
{Hoyle} F.,  {Outram} P.~J.,  {Shanks} T.,  {Boyle} B.~J.,  {Croom} S.~M.,
  {Smith} R.~J.,  2002, \mnras, 332, 311

\bibitem[\protect\citeauthoryear{{Jannuzi} \& {Dey}}{{Jannuzi} \&
  {Dey}}{1999}]{JD99}
{Jannuzi} B.~T.,  {Dey} A.,  1999, in {Weymann} R.,  et~al. eds, ASP Conf. Ser.
  191: Photometric Redshifts and the Detection of High Redshift Galaxies p.~111

\bibitem[\protect\citeauthoryear{{Kaiser}}{{Kaiser}}{1987}]{Kaiser87}
{Kaiser} N.,  1987, \mnras, 227, 1

\bibitem[\protect\citeauthoryear{{Landy} \& {Szalay}}{{Landy} \&
  {Szalay}}{1993}]{LS93}
{Landy} S.~D.,  {Szalay} A.~S.,  1993, \apj, 412, 64

\bibitem[\protect\citeauthoryear{{Le F{\`e}vre} et~al.,}{{Le F{\`e}vre}
  et~al.}{2005}]{LeFevre05}
{Le F{\`e}vre} O.,  et~al., 2005, \aap, 439, 877

\bibitem[\protect\citeauthoryear{{Lewis} et~al.,}{{Lewis}
  et~al.}{2002}]{Lewis02}
{Lewis} I.~J.,  et~al., 2002, \mnras, 333, 279

\bibitem[\protect\citeauthoryear{{Li}, {Kauffmann}, {Jing}, {White},
  {B{\"o}rner} \& {Cheng}}{{Li} et~al.}{2006}]{Li06a}
{Li} C.,  {Kauffmann} G.,  {Jing} Y.~P.,  {White} S.~D.~M.,  {B{\"o}rner} G.,
   {Cheng} F.~Z.,  2006, \mnras, 368, 21

\bibitem[\protect\citeauthoryear{{Loveday}, {Peterson}, {Maddox} \&
  {Efstathiou}}{{Loveday} et~al.}{1996}]{Loveday96}
{Loveday} J.,  {Peterson} B.~A.,  {Maddox} S.~J.,    {Efstathiou} G.,  1996,
  \apjs, 107, 201

\bibitem[\protect\citeauthoryear{{Madgwick} et~al.,}{{Madgwick}
  et~al.}{2003}]{Madgwick03}
{Madgwick} D.~S.,  et~al., 2003, \mnras, 344, 847

\bibitem[\protect\citeauthoryear{{Mart{\'{\i}}nez} \& {Saar}}{{Mart{\'{\i}}nez}
  \& {Saar}}{2002}]{Martinez02book}
{Mart{\'{\i}}nez} V.~J.,  {Saar} E.,  2002, Statistics of the Galaxy
  Distribution.
Chapman \& Hall/CRC

\bibitem[\protect\citeauthoryear{{Matsubara} \& {Suto}}{{Matsubara} \&
  {Suto}}{1996}]{Matsubara96}
{Matsubara} T.,  {Suto} Y.,  1996, \apjl, 470, L1+

\bibitem[\protect\citeauthoryear{{Matsubara} \& {Szalay}}{{Matsubara} \&
  {Szalay}}{2001}]{Matsubara01}
{Matsubara} T.,  {Szalay} A.~S.,  2001, \apjl, 556, L67

\bibitem[\protect\citeauthoryear{{Metcalfe}, {Shanks}, {Campos}, {McCracken} \&
  {Fong}}{{Metcalfe} et~al.}{2001}]{Metcalfe01}
{Metcalfe} N.,  {Shanks} T.,  {Campos} A.,  {McCracken} H.~J.,    {Fong} R.,
  2001, \mnras, 323, 795

\bibitem[\protect\citeauthoryear{{Norberg} et~al.,}{{Norberg}
  et~al.}{2002}]{Norberg02a}
{Norberg} P.,  et~al., 2002, \mnras, 332, 827

\bibitem[\protect\citeauthoryear{{Peacock} et~al.,}{{Peacock}
  et~al.}{2001}]{Peacock01}
{Peacock} J.~A.,  et~al., 2001, \nat, 410, 169

\bibitem[\protect\citeauthoryear{{Peebles}}{{Peebles}}{1980}]{Peebles80}
{Peebles} P.~J.~E.,  1980, {The Large-Scale Structure of the Universe}.
Princeton University Press.

\bibitem[\protect\citeauthoryear{{Peebles}}{{Peebles}}{1984}]{Peebles84}
{Peebles} P.~J.~E.,  1984, \apj, 284, 439

\bibitem[\protect\citeauthoryear{{Percival} et~al.,}{{Percival}
  et~al.}{2002}]{Percival02}
{Percival} W.~J.,  et~al., 2002, \mnras, 337, 1068

\bibitem[\protect\citeauthoryear{{Percival} et~al.,}{{Percival}
  et~al.}{2006a}]{Percival06a}
{Percival} W.~J.,  et~al., 2006a, astro-ph/0608635

\bibitem[\protect\citeauthoryear{{Percival} et~al.,}{{Percival}
  et~al.}{2006b}]{Percival06b}
{Percival} W.~J.,  et~al., 2006b, astro-ph/0608636

\bibitem[\protect\citeauthoryear{{Phillipps}, {Fong}, {Fall} \&
  {MacGillivray}}{{Phillipps} et~al.}{1978}]{Phillipps78}
{Phillipps} S.,  {Fong} R.,  {Fall} R.~S.~E.~S.~M.,    {MacGillivray} H.~T.,
  1978, \mnras, 182, 673

\bibitem[\protect\citeauthoryear{{Phleps}, {Peacock}, {Meisenheimer} \&
  {Wolf}}{{Phleps} et~al.}{2006}]{Phleps06}
{Phleps} S.,  {Peacock} J.~A.,  {Meisenheimer} K.,    {Wolf} C.,  2006, \aap,
  457, 145

\bibitem[\protect\citeauthoryear{{Press}, {Teukolsky}, {Vetterling} \&
  {Flannery}}{{Press} et~al.}{1992}]{Press92}
{Press} W.~H.,  {Teukolsky} S.~A.,  {Vetterling} W.~T.,    {Flannery} B.~P.,
  1992, {Numerical Recipes in FORTRAN: The Art of Scientific Computing}.
Cambridge University Press.

\bibitem[\protect\citeauthoryear{{Ratcliffe}, {Shanks}, {Parker} \&
  {Fong}}{{Ratcliffe} et~al.}{1998}]{Ratcliffe98c}
{Ratcliffe} A.,  {Shanks} T.,  {Parker} Q.~A.,    {Fong} R.,  1998, \mnras,
  296, 191

\bibitem[\protect\citeauthoryear{{Roseboom} et~al.,}{{Roseboom}
  et~al.}{2006}]{Roseboom06}
{Roseboom} I.~G.,  et~al., 2006, pre-print, (astro-ph/0609178)

\bibitem[\protect\citeauthoryear{{S{\'a}nchez}, {Baugh}, {Percival}, {Peacock},
  {Padilla}, {Cole}, {Frenk} \& {Norberg}}{{S{\'a}nchez}
  et~al.}{2006}]{Sanchez06}
{S{\'a}nchez} A.~G.,  {Baugh} C.~M.,  {Percival} W.~J.,  {Peacock} J.~A.,
  {Padilla} N.~D.,  {Cole} S.,  {Frenk} C.~S.,    {Norberg} P.,  2006, \mnras,
  366, 189

\bibitem[\protect\citeauthoryear{{Saunders}, {Rowan-Robinson} \&
  {Lawrence}}{{Saunders} et~al.}{1992}]{Saunders92}
{Saunders} W.,  {Rowan-Robinson} M.,    {Lawrence} A.,  1992, \mnras, 258, 134

\bibitem[\protect\citeauthoryear{{Schulz} \& {White}}{{Schulz} \&
  {White}}{2006}]{Schulz06}
{Schulz} A.~E.,  {White} M.,  2006, Astroparticle Physics, 25, 172

\bibitem[\protect\citeauthoryear{{Scranton} et~al.,}{{Scranton}
  et~al.}{2002}]{Scranton02}
{Scranton} R.,  et~al., 2002, \apj, 579, 48

\bibitem[\protect\citeauthoryear{{Shanks}, {Bean}, {Ellis}, {Fong},
  {Efstathiou} \& {Peterson}}{{Shanks} et~al.}{1983}]{Shanks83b}
{Shanks} T.,  {Bean} A.~J.,  {Ellis} R.~S.,  {Fong} R.,  {Efstathiou} G.,
  {Peterson} B.~A.,  1983, \apj, 274, 529

\bibitem[\protect\citeauthoryear{{Smith}, {Scoccimarro} \& {Sheth}}{{Smith}
  et~al.}{2007}]{Smith07}
{Smith} R.~E.,  {Scoccimarro} R.,    {Sheth} R.~K.,  2007, \prd, 75, 063512

\bibitem[\protect\citeauthoryear{{Spergel} et~al.,}{{Spergel}
  et~al.}{2003}]{Spergel03}
{Spergel} D.~N.,  et~al., 2003, \apjs, 148, 175

\bibitem[\protect\citeauthoryear{{Spergel} et~al.,}{{Spergel}
  et~al.}{2006}]{Spergel06}
{Spergel} D.~N.,  et~al., 2006, astro-ph/0603449

\bibitem[\protect\citeauthoryear{{Tegmark} et~al.,}{{Tegmark}
  et~al.}{2006}]{Tegmark06}
{Tegmark} M.,  et~al., 2006, \prd, 74, 123507

\bibitem[\protect\citeauthoryear{{Wake} et~al.,}{{Wake}  et~al.}{2006}]{Wake06}
{Wake} D.~A.,  et~al., 2006, \mnras, 372, 537

\bibitem[\protect\citeauthoryear{{White}, {Zheng}, {Brown}, {Dey} \&
  {Jannuzi}}{{White} et~al.}{2007}]{White07}
{White} M.,  {Zheng} Z.,  {Brown} M.~J.~I.,  {Dey} A.,    {Jannuzi} B.~T.,
  2007, \apjl, 655, L69

\bibitem[\protect\citeauthoryear{{Wolf}, {Dye}, {Kleinheinrich},
  {Meisenheimer}, {Rix} \& {Wisotzki}}{{Wolf} et~al.}{2001}]{Wolf01}
{Wolf} C.,  {Dye} S.,  {Kleinheinrich} M.,  {Meisenheimer} K.,  {Rix} H.-W.,
  {Wisotzki} L.,  2001, \aap, 377, 442

\bibitem[\protect\citeauthoryear{{York} et~al.,}{{York}  et~al.}{2000}]{York00}
{York} D.~G.,  et~al., 2000, \aj, 120, 1579

\bibitem[\protect\citeauthoryear{{Zehavi}, {Blanton}, {Frieman}, {Weinberg},
  {Waddell}, {Yanny} \& {York}}{{Zehavi} et~al.}{2002}]{Zehavi02}
{Zehavi} I.,  {Blanton} M.~R.,  {Frieman} J.~A.,  {Weinberg} D.~H.,  {Waddell}
  P.,  {Yanny} B.,    {York} D.~G.,  2002, \apj, 571, 172

\bibitem[\protect\citeauthoryear{{Zehavi} et~al.,}{{Zehavi}
  et~al.}{2004}]{Zehavi04}
{Zehavi} I.,  et~al., 2004, \apj, 608, 16

\bibitem[\protect\citeauthoryear{{Zehavi} et~al.,}{{Zehavi}
  et~al.}{2005}]{Zehavi05a}
{Zehavi} I.,  et~al., 2005, \apj, 621, 22

\end{thebibliography}

\end{document}